\documentclass[fleqn,usenatbib]{mnras}

\usepackage{newtxtext,newtxmath}

\usepackage[T1]{fontenc}

\DeclareRobustCommand{\VAN}[3]{#2}
\let\VANthebibliography\thebibliography
\def\thebibliography{\DeclareRobustCommand{\VAN}[3]{##3}\VANthebibliography}
\usepackage{bm}

\usepackage{graphicx}	
\usepackage{amsmath}	
\usepackage{algorithm}
\usepackage{algpseudocode}
\usepackage{caption}
\usepackage{lineno}
\usepackage{subcaption}
\usepackage{multicol}
\usepackage{paralist}
\usepackage{natbib}
\usepackage{array}
\usepackage{booktabs}
\usepackage{diagbox}
\usepackage{makecell}

\usepackage{xcolor}
\definecolor{pink}{rgb}{0.96, 0.76, 0.76}
\definecolor{aqua}{rgb}{0.22, 0.96, 0.93}
\definecolor{darkred}{rgb}{0.76, 0.23, 0.13}

\title[The rotation curve of the Milky Way]{The rotation curve of the Milky Way measured by Classical Cepheids from \emph{Gaia} DR3}

\author[Feng et al.]{Qikang Feng$^{1, 2}$, Yang Huang$^{3, 4}$\thanks{E-mail: huangyang@ucas.ac.cn}, Huawei Zhang$^{1, 2}$\thanks{E-mail: zhanghw@pku.edu.cn}, Jifeng Liu$^{4, 3}$
\\
$^{1}$Department of Astronomy, School of Physics, Peking University, Beijing, 100871, China\\
$^{2}$Kavli Institute for Astronomy and Astrophysics, Peking University, Beijing, 100871, China\\
$^{3}$chool of Astronomy and Space Science, University of Chinese
Academy of Sciences, Beijing, 100049, China\\
$^{4}$Key Laboratory of Optical Astronomy, National Astronomical Observatories, Chinese Academy of Science Beijing, 100012, China
}

\date{Accepted XXX. Received YYY; in original form ZZZ}

\pubyear{\the\year{}}
\begin{document}
\label{firstpage}
\pagerange{\pageref{firstpage}--\pageref{lastpage}}
\maketitle

\begin{abstract}
We determine the rotation curve (RC) of the Milky Way in the range of $6<R<18\ \rm{kpc}$ using a sample of 903 carefully selected classical Cepheids with precise proper motions and high-quality radial velocities from \emph{Gaia} DR3.
Their distances can be accurately measured from the well-known Period–Wesenheit relations. 
The RC is computed from the three-dimensional velocities components of these Cepheids. 
Generally, the RC shows a slight decline with distance from the Galactic center. 
On top of this general trend, the newly constructed RC shows a dip around $R\sim 10$--$11\ \rm{kpc}$, followed by a bump around $R\sim 13$--$14\ \rm{kpc}$. 
This feature has also been reported in other RC measurements, mostly in RCs traced by young tracers like Cepheids. To better constrain the Milky Way mass, an averaged RC is then constructed by combining measurements from this work and previous efforts. Due to the ambiguous nature of the dip-and-bump feature, this averaged RC is constructed only within the radial range where the RC appears to be less influenced by this feature.
By using this averaged RC, we determine the circular velocity at solar position and also build a parameterized mass model of our Milky Way. The result for the circular velocity at the solar position is $V_{\mathrm{c}}(R_{0})=236.8\pm0.8\ \rm{km\ s^{-1}}$, which is in good agreement with previous measurements. The local dark matter density and the enclosed dark matter halo mass within 18 kpc are estimated from the averaged RC under different baryonic models, yielding a series of consistent results: a local density of $0.33$--$0.40\ \mathrm{GeV\ cm^{-3}}$ and an enclosed mass of $1.19$--$1.45\times10^{11}\ M_{\odot}$.
\end{abstract}




\section{Introduction}\label{intro}
The rotation curve (hereafter RC) describes the circular speed at different radii from the center of galaxies. 
It is a powerful tool to constrain the mass distribution of the galaxies \citep{VD59, vA85}, which is one of the most fundamental properties for understanding the galaxy formation and evolution \citep{LA89, DB94, LX12, DE19}. 
Particularly, the RC provides an advantageous window to investigate the distribution of the unseen dark matter \citep[hereafter DM;][]{VD59, FR70} whose existence is exactly disclosed from the measurement of the RC \citep{RB80}. 
Specifically, the local DM density \citep{SA10, WB10}, which can be constrained by the RC, could provide crucial constraints for ground- and space-based experiments aimed at detecting DM particles.
\par{}
The RC of our Milky Way has been measured by many attempts with a variety of tracers. 
For example, the RC of the inner Galactic disc within the solar orbit can be measured by the so-called tangent-point method, which takes the gaseous tracers like H\uppercase\expandafter{\romannumeral 1} 21 cm or CO 2.6 mm gas emissions in the Galactic plane \citep{BG78, GU79, FI89, LE08, SO09}. 
For the RC in regions beyond the solar orbit, including the outer disc and the halo, the tangent-point method is no longer applicable. In these regions, the RC must be determined using three-dimensional velocity components from various tracers, which require accurate distance measurements, proper motions, and line-of-sight velocities.
Several commonly used tracers for measuring the RC in the outer disc region include classical Cepheids \citep{PT97, MO19, KW19, AB20}, red clumps \citep{BV12b, HU16}, OB stars \citep{BB15}, carbon stars \citep{BT13}, H \uppercase\expandafter{\romannumeral 2} regions \citep{BB93} and masers \citep{RE14}, while the RC of the halo region is usually derived by the velocity anisotropy parameter of the distant tracers including blue horizontal brunch stars \citep{XU08, KF12}, stellar clusters \citep{HR87, SF09} and dwarf galaxies \citep{BG05, SO091}.

\par{}
Building a large sample of stars with precise distances and complete kinematic data, such as proper motions and radial velocities, has long been a challenge.  These limitations introduce considerable uncertainties in determining the RC profile, particularly in the outer regions beyond the solar orbit.
Fortunately, this situation has been improved significantly thanks to the powerful astrometric mission \emph{Gaia}, which provides accurate positions and proper motions for about few billions of Milky Way stars, as well as radial velocities for tens of millions of  stars \citep{GC16, GC18}. In addition, spectroscopic data not only provide radial velocity measurements but also allow for the derivation of more precise distances (e.g., \citealt{HG19, ZH23}). Such data are made available through large-scale spectroscopic surveys like APOGEE \citep{MJ17} and LAMOST \citep{LU15}.
By combining data from these surveys, it has become easier to obtain reliable distances and kinematic information for a large number of stellar tracers. Taking advantage of these data, many studies have provided new insights into the Milky Way’s RC, particularly in the outer regions. For instance, several studies have reported a sharp decline in the RC beyond 15--20 kpc \citep[e.g.,][]{WH23, JY23}, suggesting that the DM mass in the Milky Way may be significantly lower than previously estimated. However, other works have suggested that this apparent decline could stem from limitations inherent in the framework that is commonly adopted in previous RC studies, specifically, the systematic biases associated with using the Jeans equation for asymmetric drift corrections and the breakdown of the axisymmetric assumption, especially in the outer disc \citep{OU25, KA25}. Therefore the validity of this framework needs to be further examined.
\par{}
In this work, we measure the Milky Way's RC using classical Cepheids as tracers. Cephieds are excellent tracers for measuring the RC for three main reasons. First, their distances can be accurately determined using the well-established period-luminosity relation. Compared to other tracers typically used to study the kinematics of the thin disc, such as Red Giant Branch (RGB) stars \citep{EI19, ZH23}, Cepheids provide a considerably more accurate determination of distances. Second, Cepheids are primarily located in the thin disc, which makes them less likely to be contaminated by stars from the thick disc or halo populations. Third, Cepheids are relatively young, meaning they are less affected by dynamical heating. This ensures that the RC traced by Cepheids is minimally influenced by the asymmetric drift (see Section 3 for details). These advantages, especially their minimal asymmetric drift, make the RC measurement based on Cepheids a valuable tool for cross-checking RCs derived from other tracers using the framework adopted in previous studies. Since \cite{AB20}, several new and larger Cepheid datasets with updated kinematic information have become available. Therefore, it is important to remeasure and reanalyze the Milky Way RC using these new data. Here we select the classical Cepheids sample from \emph{Gaia} Data Release 3 \citep[DR3;][]{RI23}. In the latest data release, \emph{Gaia} has provided a very large Cepheids catalog, with approximately 1.5 times more stars compared to Data Release 2 \citep{CL19}. This expanded catalog provides a larger sample for measuring the RC across a broad range of the Galaxy.

\section{Sample selection}
We select the sample stars classified as ``DCEP" from the \emph{Gaia} DR3 Cepheid catalog \citep{RI23}, and exclude the member stars of the Large Magellanic Cloud (LMC), Small Magellanic Cloud (SMC), M31 and M33 galaxies according to their positions on $l-b$ diagram. Following the approach of other studies using \emph{Gaia} data, we select only the sample stars with a \emph{Gaia} renormalized unit weight error (RUWE) less than 1.4 to ensure good astrometric quality. We also select only the stars with radial velocity measurements, as all three velocity components of the tracers will be used in this study.

\subsection{Radial velocity adoption}
In \emph{Gaia} DR3 catalog, the radial velocity measurements from two methods are given for those Cepheids. The first method is the `model method' which derives the radial velocity results from radial velocity curves \citep{BU93}. Due to the pulsating nature of Cepheids, radial velocity measurements taken at different epochs exhibit variability. 
Therefore, accurate determination of a Cepheid's radial velocity requires precise corrections for pulsational velocity variations at different epochs \citep{CL23}, typically based on a well-established model of the pulsating radial velocity curve. The second method, on the other hand, is the `combined method', which gives the radial velocity results by combining several measurements from different epochs without accounting for the pulsating radial velocity curve.
Although the `model method' provides more accurate results, the number of Cepheids with radial velocities obtained through this method in \emph{Gaia} DR3 catalog is very limited. To construct a larger sample, we employ the radial velocities provided by the `combined method'.

In fact, when the number of epochs is sufficiently high, the results from the combined method closely match those obtained from the `model method'. In the \emph{Gaia} sample, the `model method' is applicable only for Cepheids with at least seven epochs. 
By comparing the radial velocities from the `combined method' to those from the `model method', we find that the scatter is about 11 $\mathrm{km\ s^{-1}}$ when the number of epochs is seven. This scatter tends to significantly reduce when the number of epochs is eight or more, decreasing to $6.1\ \mathrm{km\ s^{-1}}$ with a minimal offset of $0.4\ \mathrm{km\ s^{-1}}$.
To balance the accuracy of radial velocity and the sample size, we adopt the classical Cepheids with eight or more epochs in our sample.
In total, 1986 stars are left.

\subsection{6D positions and velocities}
\par{}
Distances of these Cepheids are derived by the apparent and absolute Wesenheit magnitudes \citep{RI19}, which are computed by the apparent intensity averaged magnitude of the three bands: $G$, $G_{BP}$, $G_{RP}$ and the Period-Wesenheit (PW) relations for \emph{Gaia} Cepheids \citep{RI23}, respectively. This magnitude is defined to be reddening-free; thus we do not need any reddening corrections.  The typical uncertainty in the distance determinations for our sample is approximately 7\%, as assessed using open clusters \citep{H24}. This uncertainty value is smaller than the typical uncertainty associated with tracers used to measure the RC or perform other kinematic studies, such as the RGB sample used in \cite{EI19} and \cite{ZH23} (around 10\% when only accounting for the calibration uncertainties). It is worth noting that \citet[][hereafter SK25]{SK25} recently pointed out that the reddening-free assumption of the Wesenheit magnitude may introduce additional uncertainties. By comparing our distances with those from SK25 (see Figure A1), we find a median offset of 6.6\% and a scatter of 13.6\%. Assuming that both measurements contribute equally, the distance uncertainty of our measurements inferred from this scatter is consistent with that derived from the open-cluster test, indicating that the latter is not underestimated. Moreover, although slight differences exist between our distances and those from SK25, we find that such small discrepancies in distance estimates do not significantly affect the RC measurements, as confirmed by repeating the RC measurement using the distances from SK25 (see Appendix).

Using radial velocities and proper motions from \emph{Gaia}, the 3D positions and 3D velocities of these Cepheids can be determined. During the calculations, the Sun is fixed at $(R_{\odot},\ \phi_{\odot},\ Z_{\odot})\ =\ (8.275,\,0,\,0.025)$\,kpc \citep{GR21, JU08}. 
The Galactocentric velocity of the Sun is adopted as $V_{R,\odot}=11.1\ \rm{km\ s^{-1}}$ \citep{SH10}, $V_{\phi, \odot}=250.2\ \rm{km\ s^{-1}}$ and $V_{Z,\odot}=7.9\ \rm{km\ s^{-1}}$, which is derived from the proper motion of Sgr~A$^*$ \citep{RE04}.
We also examine the results of this study using alternative values for the Sun's peculiar velocities \citep{HU15, WF21}, and the main findings remain consistent (See section 3.3).
\par{}
After deriving the 3D positions and 3D velocities of the Cepheids, we select the sample stars within the range of $R > 6\ \mathrm{kpc}$ and within $\pm30^{^\circ}$ of the anti-Galactic center direction to derive the RC. Also, outliers with $|Z|>0.5\ \rm{kpc}$ or $|V_{Z}|>100\ \rm{km\ s^{-1}}$ are removed. Among these selections, the wedge near the anti-Galactic center direction is chosen for two main reasons. First, for stars located in this region, radial-velocity uncertainties, which dominate the errors in the stellar three-dimensional velocity measurements, have a smaller effect on the determination of the Galactocentric rotational velocity ($V_{\phi}$). Second, the geometric effect of the disc warp is minimized within this wedge, as the line-of-nodes of the warp is oriented close to the anti-Galactic center direction \citep{H24, HV25}; consequently, this wedge better aligns with the axisymmetric assumption adopted in our subsequent analysis (see the next section). A similar wedge has also been employed in previous studies \citep[e.g.,][]{EI19, ZH23, OU24}.
After applying these criteria, a total of 903 Cepheids are left. The spatial distributions of our final sample are illustrated in Figure~1.

\begin{figure*}
    \begin{center}
    \includegraphics[width=0.9\textwidth]{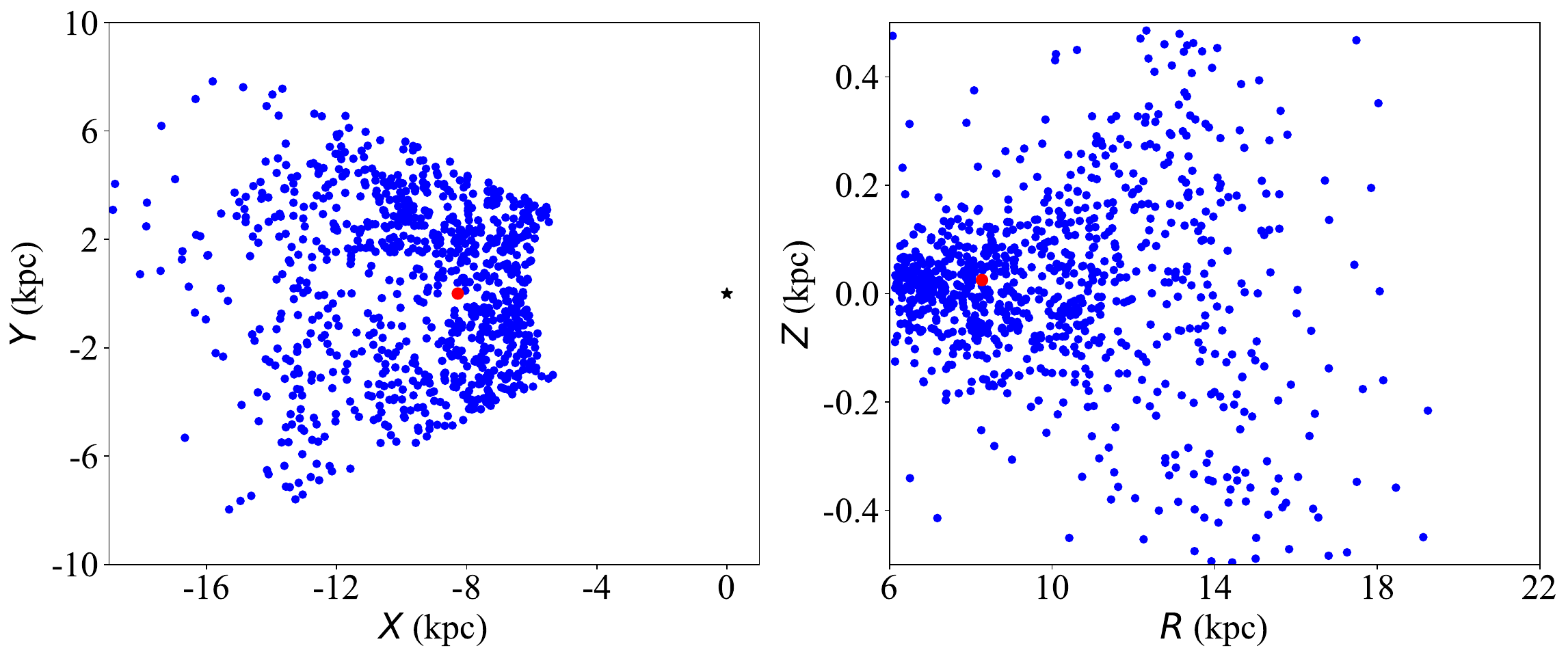}
    \end{center}
    \caption{Left panel: The $X-Y$ distribution of the final adopted sample. The black star denotes the position of the Galactic center. Right panel: the $R-Z$ distribution of the final adopted sample. The position of the Sun is marked by the red dot in both panels.} 
    \label{Fig1}
\end{figure*}

\begin{figure}
    \begin{center}
    \includegraphics[width=0.48\textwidth]{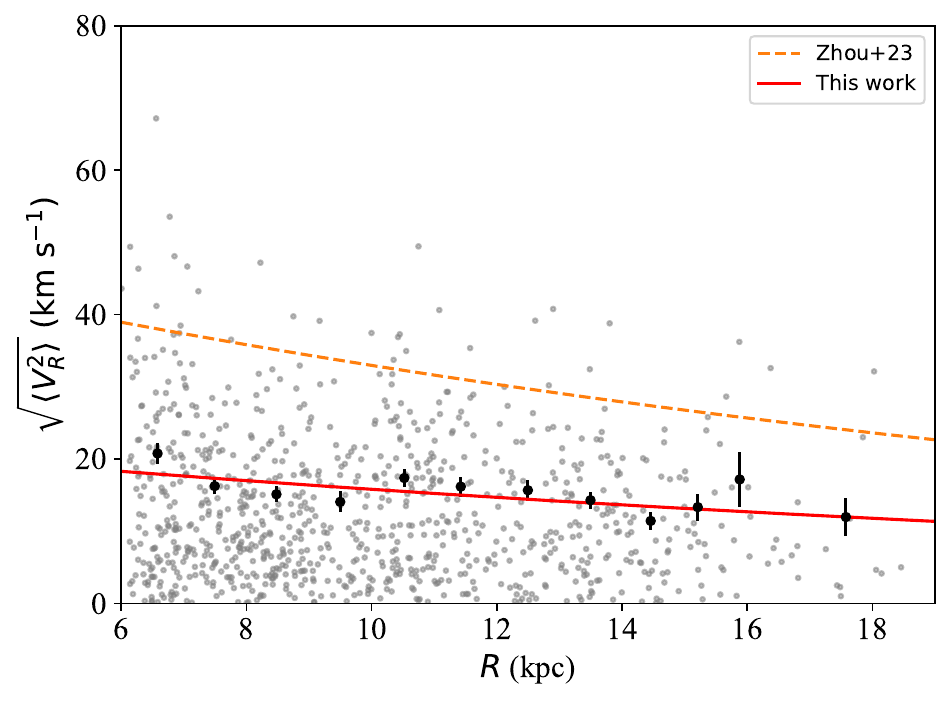}
    \end{center}
    \caption{Radial profiles of the radial velocity tensor $\sqrt{\langle V_{R}^2 \rangle}$. Grey dots indicate measurements for individual stars. The black dots represent the square root of the $\langle V_{R}^2 \rangle$ in each bin, with uncertainties calculated via bootstrapping with 100 resamples. The red line represents the best-fit exponential function to our sample’s $\sqrt{\langle V_{R}^2 \rangle}$, while the orange dashed line corresponds to the best-fit exponential model for $\sqrt{\langle V_{R}^2 \rangle}$ from \protect\cite{ZH23}. Overall, our derived $\sqrt{\langle V_{R}^2 \rangle}$ values are much lower than those reported in \protect\cite{ZH23} across the radial range.}
    \label{Fig2}
\end{figure}

\section{Rotation Curve}
\subsection{Asymmetric drift}
In principle, the RC is not the same as the mean azimuthal velocity $\langle V_{\phi} \rangle$. The difference between RC and $\langle V_{\phi} \rangle$ is known as the asymmetric drift. This difference should be considered when determining the RC, especially for old tracers. In the framework widely adopted in previous studies, the asymmetric drift is accounted for and corrected using the Jeans equation. The details are as follows:
Assuming an axisymmetric mass distribution for the Milky Way, its RC is derived by:
\begin{equation}
    V_{\mathrm{c}}^{2}(R)=R\frac{\partial\Phi}{\partial R}\ (\mathrm{at}\ z=0).
\end{equation}
Combined with the Jeans Equation \citep{BT87}, the RC can be calculated by the velocity tensors of the tracers:
\begin{equation}
    V_{\mathrm{c}}^{2}(R)=\langle V_{\phi}^2 \rangle-\langle V_{R}^2 \rangle(1-\frac{R}{R_{d}}-\frac{2R}{R_{\sigma}})
\end{equation}
where $\langle V_{\phi}^2 \rangle$ and $\langle V_{R}^2 \rangle$ are the mean values of the azimuthal and radial velocity tensors, calculated from an ensemble of stars within each radial bin. $R_{d}$ and $R_{\sigma}$ represent, respectively, the scale length of the exponential tracer density profile and the scale length of the exponential radial velocity dispersion profile. The term involving $\frac{\partial \overline{V_{R}V_{Z}}}{\partial Z}$ has already been neglected in this equation, as the covariance $\overline{V_{R}V_{Z}}$ shows no significant variation for sample stars located near the disc plane \citep{BV15}. 

Since Cepheids are both young and kinematically cold, previous RC measurements using Cepheids as tracers have neglected the asymmetric drift, as they are very small \citep{KW19, AB20}. However, here we not only carefully verify whether the asymmetric drift is indeed small by estimating it using the Jeans equation, but also further determine the RC by explicitly accounting for this effect (i.e., by deriving it through Equation~2 rather than simply adopting $\langle V_{\phi} \rangle$), even for the young Cepheid sample, in order to obtain a more accurate result.

In general, the asymmetric drift increases with the radial velocity dispersion \(\langle V_{R}^2 \rangle\). To estimate its impact, we present \(\sqrt{\langle V_{R}^2 \rangle}\) for our sample and compare it with that of the older RGB sample from \cite{ZH23} (see Figure 2), providing a rough estimate of the asymmetric drift in our dataset. From Figure 2, we find that the $\sqrt{\langle V_{R}^2 \rangle}$ of our Cepheid sample is only about half that of the RGB sample in \cite{ZH23}, clearly indicating that Cepheids are kinematically colder and experience significantly less influence from asymmetric drift compared to older tracers.

Besides $\langle V_{R}^2 \rangle$, accurately determining the scale length of the tracer density profile, $R_{d}$, and the scale length of the radial velocity dispersion profile, $R_{\sigma}$, is also essential. As in other RC measurement studies \citep[e.g.,][]{HU16, EI19, ZH23, OU24}, we adopt the value of $R_{d}$ from the literature and determine $R_{\sigma}$ by fitting the $\sqrt{\langle V_{R}^2 \rangle}-R$ profile of our sample stars. 
Numerous studies estimating $R_{d}$ for the Milky Way's thin disc using older stellar tracers have reached near consensus \citep{BJ05, BV12a, BH16, CH17}. However, the scale length of the young disc remains poorly constrained. Observations suggest that it is longer than that of the old thin disc \citep{WN17, MC17, XI18}, but its precise value remains uncertain. Given this, we adopt $R_{d} = 4$~kpc, based on the average of values reported in \citet{BV12a}, \citet{WN17}, and \citet{XI18}. 
The expected value for $R_{\sigma}$ is $R_{\sigma} = 2R_{d}$ \citep{VK82}, though many measurements indicate even larger values \citep{SR14, HU16, MC19, EI19, ZH23}. In this study, following the methodology of \citet{HU16, EI19, ZH23, OU24}, we derive $R_{\sigma}$ by fitting the $\langle V_{R}^2 \rangle$ profile with an exponential function. The fitting result, shown in Figure 2, yields $R_{\sigma} = 27.3$~kpc.
Based on these parameters, we compute the corrected RC while accounting for asymmetric drift. The average asymmetric drift for our Cepheid sample is estimated to be 1.6 $\mathrm{km\ s^{-1}}$, with the differences between $\langle V_{\phi} \rangle$ and $V_{\mathrm{c}}$ remaining below 1\% in most bins. This typical value of the asymmetric drift confirms that it is indeed small for Cepheids. This effect is negligible; nevertheless, the asymmetric drift correction is still applied here to derive a more accurate RC.

\subsection{Rotation curve}
The final measurement of the RC, incorporating the asymmetric drift correction (i.e., calculated via Equation 2), is presented in Figure 3 and Table 1, covering the range $6 < R < 18$~kpc, and the comparison between several previous studies is presented in Figure 5 \citep[][slight adjustment is applied to each RC to account for the variations in the adopted solar radius and velocity]{EI19, AB20, ZH23, JY23, PD23, OU24}. The RC measured here shows a slight declining trend generally. It reaches approximately $236\ \mathrm{km\ s^{-1}}$ at the solar radius and decreases to $230\ \mathrm{km\ s^{-1}}$ at $R\sim16\ \mathrm{kpc}$. After accounting for differences in the adopted solar radius and, consequently, in the solar rotational velocity across the literature, both the value and the overall declining trend of our RC are in good agreement with previous results.
In most of the radial bins, the uncertainties in the RC measurement are less than $2\ \mathrm{km\ s^{-1}}$, which is better than the measurement done by \cite{AB20} using the same type of tracer.
The uncertainties are also comparable to measurements obtained using older tracers \citep{EI19, ZH23, OU24}, which have sample sizes one to two orders of magnitude larger.
Notably, around $R \sim 17\text{–}18\ \mathrm{kpc}$, our RC, which is only marginally affected by the asymmetric drift, shows excellent agreement with previous measurements from both Cepheids and other tracers. This agreement suggests that the widely adopted Jeans equation–based framework introduces no significant systematic bias within this radial range, even for older stellar populations.
\begin{figure}
    \begin{center}
    \includegraphics[width=0.48\textwidth]{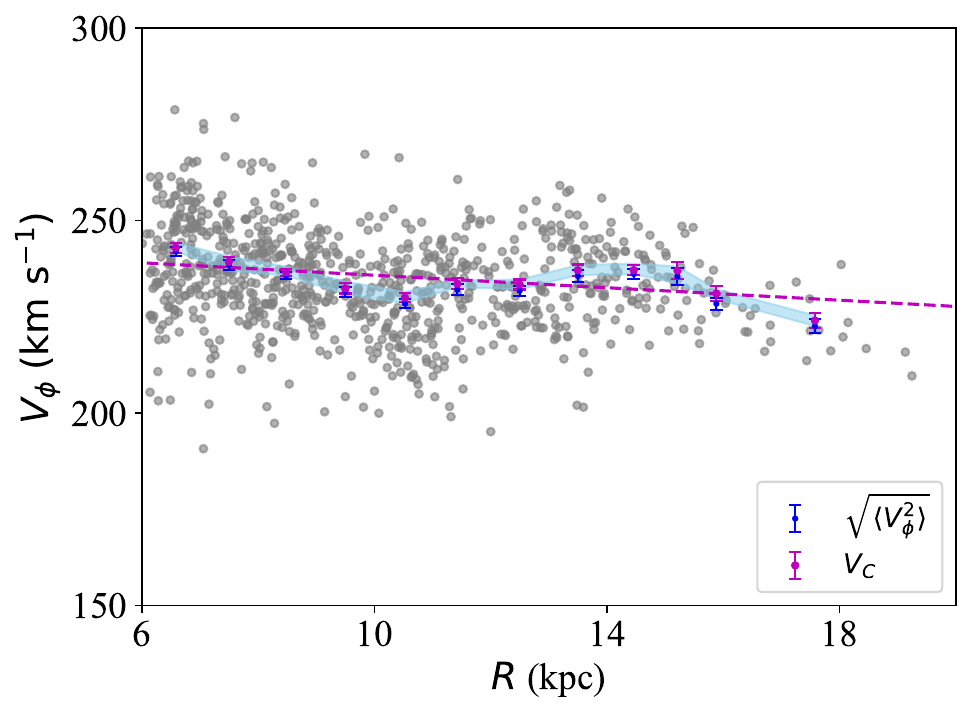}
    \end{center}
    \caption{The gray dots denote the rotation velocity $V_{\phi}$ of the sample stars. The blue dots represent the azimuthal velocity tensor, $\sqrt{\langle V_{\phi}^2 \rangle}$, calculated in radial bins. The magenta dots indicate the RC derived from Equation 2, which includes corrections for asymmetric drift. Uncertainties for both quantities (blue and magenta points) are estimated using bootstrapping with 100 resamples. The sky-blue region indicates the 1$\sigma$ uncertainty arising from the variations in $R_{0}$ and the proper motion of Sgr A* (see Section 3.3). The magenta dashed line illustrates the overall declining trend of the RC with Galactocentric radius, while localized fluctuations are superimposed on this trend.}
    \label{Fig3}
\end{figure}

\begin{table}
\footnotesize
\centering
    \caption{The values of RC measured from our Cepheids sample. The uncertainties here are calculated by bootstrapping with 100 resamples}. \label{Table1}
    \begin{tabular}{ccccccccc}
    \hline
    \emph{R}\ (kpc) & $V_{\mathrm{c}}$\ (km s$^{-1}$) & $\sigma_{V_c}$\ (km s$^{-1}$) \\
    \hline
      
    6.58 & 242.92 & 1.25\\
    7.49 & 239.31 & 1.33\\
    8.48 & 236.54 & 0.86\\
    9.50 & 232.36 & 1.32\\
    10.52 & 229.97 & 1.13\\
    11.42 & 233.57 & 1.39\\
    12.50 & 233.38 & 1.30\\
    13.50 & 237.05 & 1.53\\
    14.46 & 237.11 & 1.31\\
    15.21 & 237.00 & 2.22\\
    15.88 & 231.03 & 1.95\\
    17.58 & 224.03 & 1.91\\
    
    \hline& 
    \end{tabular}
\end{table}
Beyond the general decreasing trend, the newly derived RC also exhibits a fluctuation behavior. As shown in Figure 3, there is a dip around $R\sim10$--$11\ \mathrm{kpc}$ which is followed by a bump signal around $R\sim13$--$14\ \mathrm{kpc}$. This feature can also be observed in several previous results around similar Galactocentric radius \citep{HU16, MO19, CE19, GC21, ZO24, BE24}. More interestingly, this feature is predominantly observed in studies tracing the RC with young tracers such as Cepheids. In contrast, among studies utilizing older tracers such as RGB stars, almost none report detecting this fluctuation feature. \citep{EI19, ZH23}.

Regarding the nature of this feature, numerous studies have proposed various scenarios to explain it, yet a comprehensive understanding remains elusive. One hypothesis suggests that this feature is associated with the ``ridge line" \citep{MM19}, which refers to the diagonal ridges observed in the $V_{\phi}-R$ phase space \citep{AT18}. These ridges are likely the result of perturbations induced by the Galactic bar or spiral arms. While the model incorporating the effect of the ``ridge line" can effectively reproduce the observed RC (see Figure 3 in \cite{MM19}), this scenario struggles to explain why the feature is significantly more prominent in the RC traced by young stellar tracers, such as Cepheids. Current observational results do not indicate the absence of the ``ridge line" in the $V_{\phi}-R$ phase space for older stellar tracers \citep{WHF20, YP23}. Other explanations include the possibility that this feature is related to the 1:1 resonance of the Galactic bar \citep{DM23}, or that it arises from star formation within multiple spiral arm segments \citep{SD23}. Notably, according to the latter scenario by \cite{SD23}, this feature would dissolve rapidly, which could explain why it is most prominent when young stars are used as tracers.

\subsection{Systematic Uncertainties}
In this section, we evaluate the systematic effects on the RC from two sources.

First, we assess the systematic uncertainties associated with the adopted solar velocity. In this study, the radial component of the solar velocity ($U_{\odot}$) is taken directly from the literature. The azimuthal component, representing the sum of the circular speed at the solar position and the Sun’s peculiar velocity $V_{\odot}$, and the vertical component are both derived from the proper motion of Sgr~A$^*$ combined with the solar Galactocentric radius $R_{0}$. The impact of different $U_{\odot}$ values is tested using two alternative literature measurements \citep{HU15, WF21}, which yield no significant change in the RC. For the azimuthal and vertical components, uncertainties are propagated by accounting for the errors in both $R_{0}$ and the proper motion determinations \citep{GR21, RE04}. A Monte Carlo analysis is performed by varying $R_{0}$ and the proper motion within their $\pm 1\sigma$ uncertainties, and the resulting RC measurements show a relative systematic uncertainty below 1\%. Additionally, we manually shift the solar rotational velocity by $10\ \mathrm{km\ s^{-1}}$; even with this substantial perturbation, the RC changes by only about 4\%. Moreover, as illustrated in Figure 4 (upper panel), varying the solar velocity does not affect the overall shape of the RC. In summary, the adopted solar velocity does not significantly affect the final RC.

Second, we assess the impact on the RC of using different azimuthal wedges. The RC is remeasured using samples drawn from several distinct azimuthal ranges, and the results are shown in the lower panel of Figure 4. For comparison, we also include the case without any azimuthal selection, corresponding to the $\pm180^{\circ}$ line in the figure. The results show that the RC is not significantly affected, with the largest deviation among all tested ranges being only about 3\%.

\begin{figure}
    \begin{center}
    \includegraphics[width=0.43\textwidth]{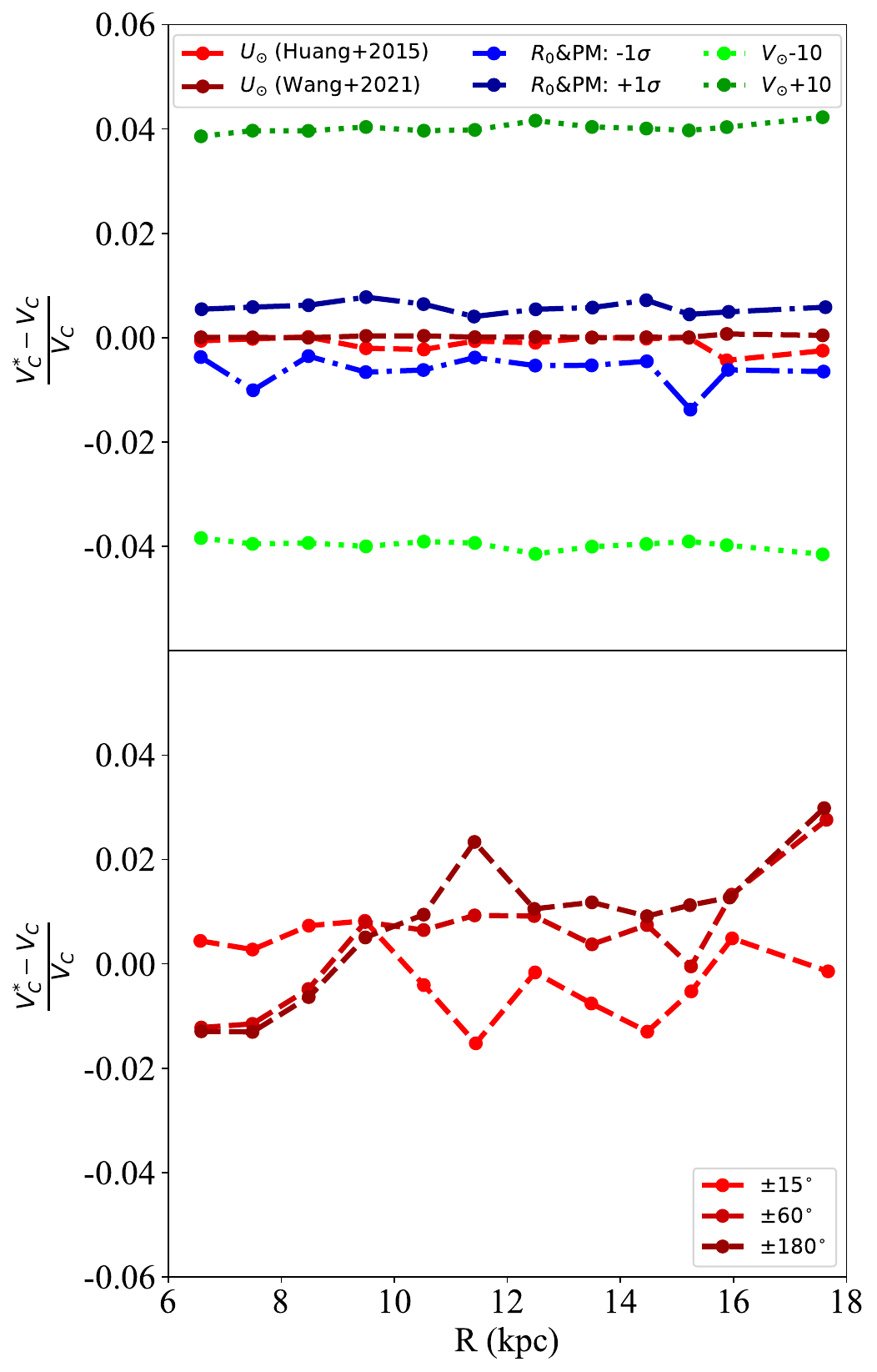}
    \end{center}
    \caption{The relative systematic uncertainties are quantified as $(V_{\mathrm{c}}^{*}- V_{\mathrm{c}})/V_{\mathrm{c}}$, where $V_{\mathrm{c}}^{*}$ represents the RC derived after accounting for different systematic effects, and $V_{\mathrm{c}}$ denotes the original measurement. The upper panel shows the systematic uncertainties in the RC arising from variations in the adopted solar velocities, whereas the lower panel illustrates those associated with different choices of azimuthal wedges.}
    \label{Fig4}
\end{figure}

\section{RC fitting and constraints on DM}
\subsection{Mass model}
\par{}
The RC is a powerful tool for constraining the mass distribution of the Milky Way, as it directly traces the Galaxy’s gravitational potential which is inherently shaped by the underlying mass distribution. Generally, the gravitational potential of the Milky Way can be described by a mass model that contains several components, including both baryonic components and DM: 
\begin{equation}
    \Phi=\sum_{i}\Phi_{i}\ ,
\end{equation}
where $\Phi_{i}$ denotes the gravitational potential of the $i^{th}$ component. Theoretically,  RC can be calculated from the sum of the contributions from those components:
\begin{equation}
    V_{\mathrm{c}}^2=\sum_{i}V_{c,\ i}^2.
\end{equation}
\par{}
Normally, the mass model of our Galaxy contains several main components: the bulge, the thin disc, the thick disc, gas (usually making a negligible contribution to the RC), and the DM halo.
Numerous previous studies have provided reliable results for the structural parameters of the baryonic components, allowing us to often fix the parameters for the first three components in the mass model. It is the mass distribution of the DM halo that remains to be constrained by the RC.

Here, in our subsequent steps to constrain the DM halo, we adopt three distinct baryonic mass models from the literature. The first model consists of a thin and a thick disc that are both described by the Miyamoto–Nagai potential \citep{MN75}, together with a bulge represented by a Plummer profile \citep{PL11}. The parameters for these components are taken from Model I of \cite{PO17}, which is also adopted in \cite{EI19} and \cite{AB20}. The second model follows the baryonic mass distribution derived by \cite{MM17}, comprising a non-spherical bulge, two exponential stellar discs, as well as HI and $\mathrm{H_{2}}$ gas discs. The third model directly adopts the baryonic components used in \cite{OU24}. In addition, we construct a composite model (hereafter referred to as “Mean”) by averaging the RC contributions from the three individual models.
To constrain the DM halo mass profile, we can use two methods. First, we can adopt the canonical Navarro-Frenk-White (NFW) profile \citep{NF95, NF96, NF97} to model the halo's mass distribution:
\begin{equation}
    \rho_{\rm{halo}}(r)=\frac{\rho_{0}}{(\frac{r}{r_{s}})(1+\frac{r}{r_{s}})^2}
\end{equation}
Here, $\rho_{s}$ and $r_{s}$ are two parameters to be estimated, representing the characteristic density of the DM halo and its scale radius, respectively. These parameters are determined by fitting the circular velocity model (which includes the fixed baryonic matter and NFW DM halo) to the RC. The enclosed mass of the DM halo within a radius $R_{\mathrm{max}}$ can then be expressed as:
\begin{equation}
    M(r<R_{\mathrm{max}})=4\pi\rho_{0}r_{s}^{3}(\ln(1+\frac{R_{\mathrm{max}}}{r_s})-\frac{R_{\mathrm{max}}}{r_s+R_{\mathrm{max}}})
\end{equation}
Alternatively, the mass profile of the DM halo can be determined directly in a model-independent way. Assuming the DM halo is spherical, the cumulative mass within a radius $R$ can be derived from the circular velocity contributed by the halo $V_{c,\ \mathrm{halo}}$:
\begin{equation}
    M(r<R)=\frac{V_{c,\ \mathrm{halo}}^{2}R}{G}
\end{equation}
Here, $V_{c,\ \mathrm{halo}}$ is obtained by subtracting the fixed baryonic contribution from the measured RC. This method provides a rough mass profile, $M(r<R)$, which is represented as discrete points corresponding to the RC measurements.

\begin{figure*}
    \begin{center}
    \includegraphics[width=0.9\textwidth]{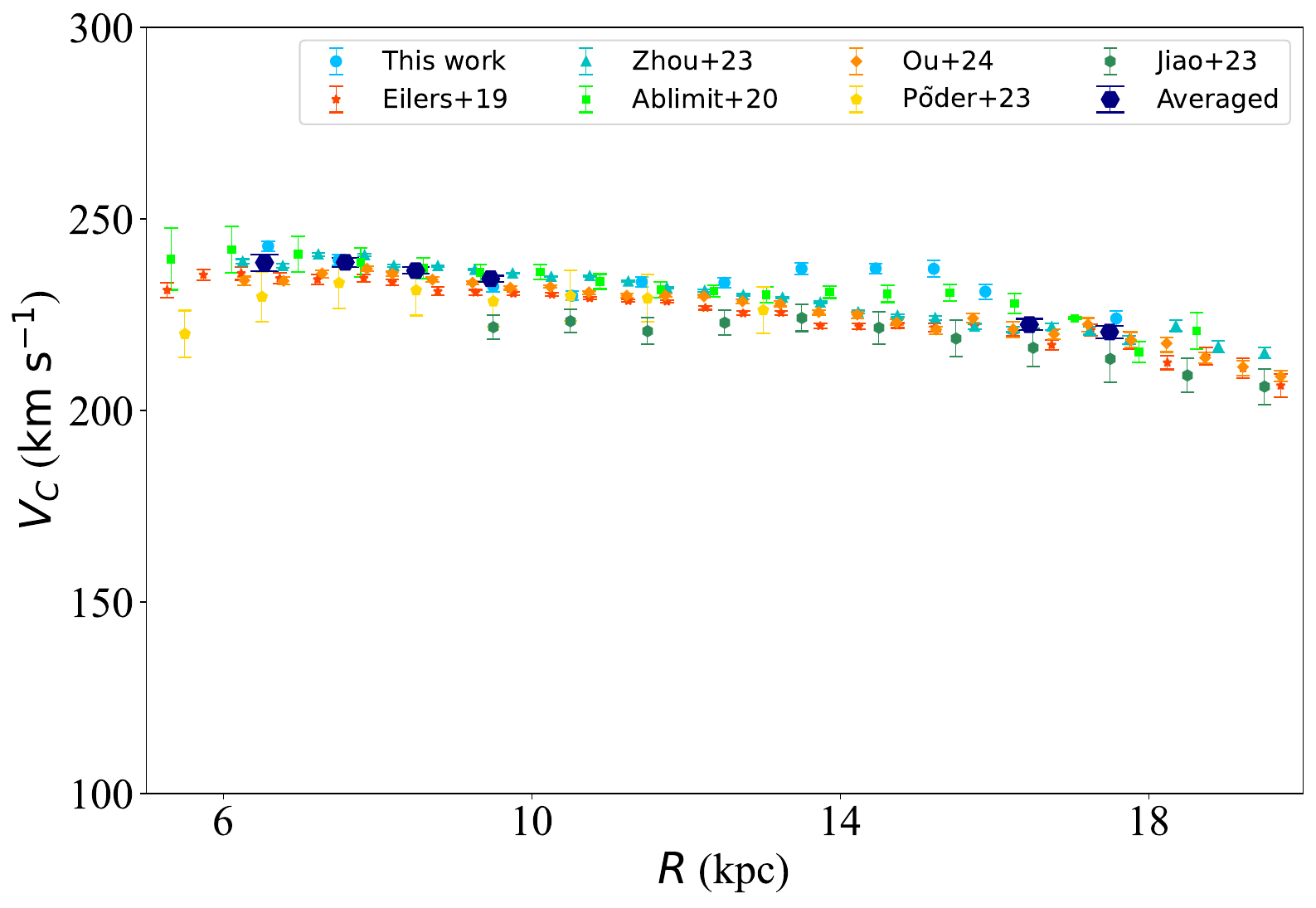}
    \end{center}
    \caption{The combined RC are denoted by the navy color hexagon points, covering only the range of $6<R<10$~kpc and $R>16$~kpc. The error bars for other points are given by the mean values of the uncertainties of the $V_{\mathrm{c}}$ given by the four measurements in each bin. The measurements by our Cepheids sample, as well as the results from previous works are also shown here by colored symbols.}
    \label{Fig4}
\end{figure*}

\subsection {Constraints on DM halo}
In estimating the DM halo mass profile, we do not directly use the measured RC results (see Figure 3 and Table 1). There are two main reasons for this. First, the RC exhibits a dip-and-bump feature, which appears to be present only in measurements traced by young samples, such as Cepheids. The underlying mechanism behind this feature is not yet well understood as described above. Second, we aim to account for systematic uncertainties that arise from using different tracers in the RC measurements. Therefore, we construct an averaged RC that spans the radial range $6<R<10$~kpc and $R>16$~kpc, where the RC appears to be less influenced by this fluctuating feature. This averaged RC is obtained by taking the mean of our own measurements and those from several previous studies \citep{AB20, ZH23, OU24} in each radial bin, with differences in the adopted solar rotational velocities among the studies taken into account.

The averaged RC is presented in Figure 5. Using this averaged RC, we first estimate the rotation speed at the solar position, yielding $V_{\mathrm{c}}(R_{\odot}) = 236.8 \pm 0.8\ \rm{km\ s^{-1}}$, which is quite consistent with previous results. Next, We apply both of the methods mentioned above to constrain the DM halo mass profile. In the NFW fitting approach, we fit the averaged RC with a mass model consisting of a fixed baryonic component and an NFW halo. The best-fit halo parameters yield the local DM density $\rho_{\odot}$ and the enclosed DM mass $M_{\mathrm{NFW}}(r<18\ \mathrm{kpc})$, where 18 kpc corresponds to the outermost radius of our RC. For the model-independent (MI) method, we derive the DM halo mass profile directly from both the measured RC in this work and the averaged RC, and determine the enclosed DM mass from the outermost radial bin, denoted as $M_{\mathrm{MI}}$. These two methods are applied to all four baryonic model choices (\citealt{PO17, MM17, OU24}, and “Mean”). The resulting DM parameters, including local densities and enclosed masses, are listed in Table 2. Figure 6 presents the inferred DM mass profiles from both the NFW fitting and the model-independent method, with the left and right panels corresponding to the three literature models and the "Mean" model, respectively. As shown in Table 2, the DM parameters obtained under different baryonic models vary only slightly: both the local DM densities and enclosed masses differ by approximately 10\% between the three literature models and the "Mean" model. Furthermore, the left panel of Figure 6 shows that the DM profiles derived from different baryonic models are closely aligned.

\begin{table*}
\centering
\caption{DM parameter results from different methods under various baryonic models.}
\begin{tabular}{ccccc}
\toprule
\diagbox[width=10em, height=3em, innerleftsep=0em, innerrightsep=0em]{Item}{Baryonic model} & \cite{PO17} & \cite{MM17} & \cite{OU24} & Mean\\
\midrule
$\rho_{\odot}\ \mathrm{(GeV\ cm^{-3})}$ & $0.33\pm0.04$ & $0.36\pm0.04$ & $0.40\pm0.04$ & $0.36\pm0.04$ \\
$M_{\mathrm{NFW}}(r<18\ \mathrm{kpc})\ (\times10^{11}\ M_{\odot})$ & $1.19\pm0.14$ & $1.31\pm0.16$ & $1.45\pm0.16$ & $1.32\pm0.15$ \\
$M_{\mathrm{MI}}$ (from measured RC) $(\times10^{11}\ M_{\odot})$ & $1.23\pm0.03$ & $1.35\pm0.03$ & $1.46\pm0.03$ & $1.35\pm0.03$ \\
$M_{\mathrm{MI}}$ (from averaged RC) $(\times10^{11}\ M_{\odot})$ & $1.16\pm0.02$ & $1.27\pm0.02$ & $1.39\pm0.02$ & $1.28\pm0.02$ \\
\bottomrule
\end{tabular}
\end{table*}

\begin{figure*}
    \begin{center}
    \includegraphics[width=1\textwidth]{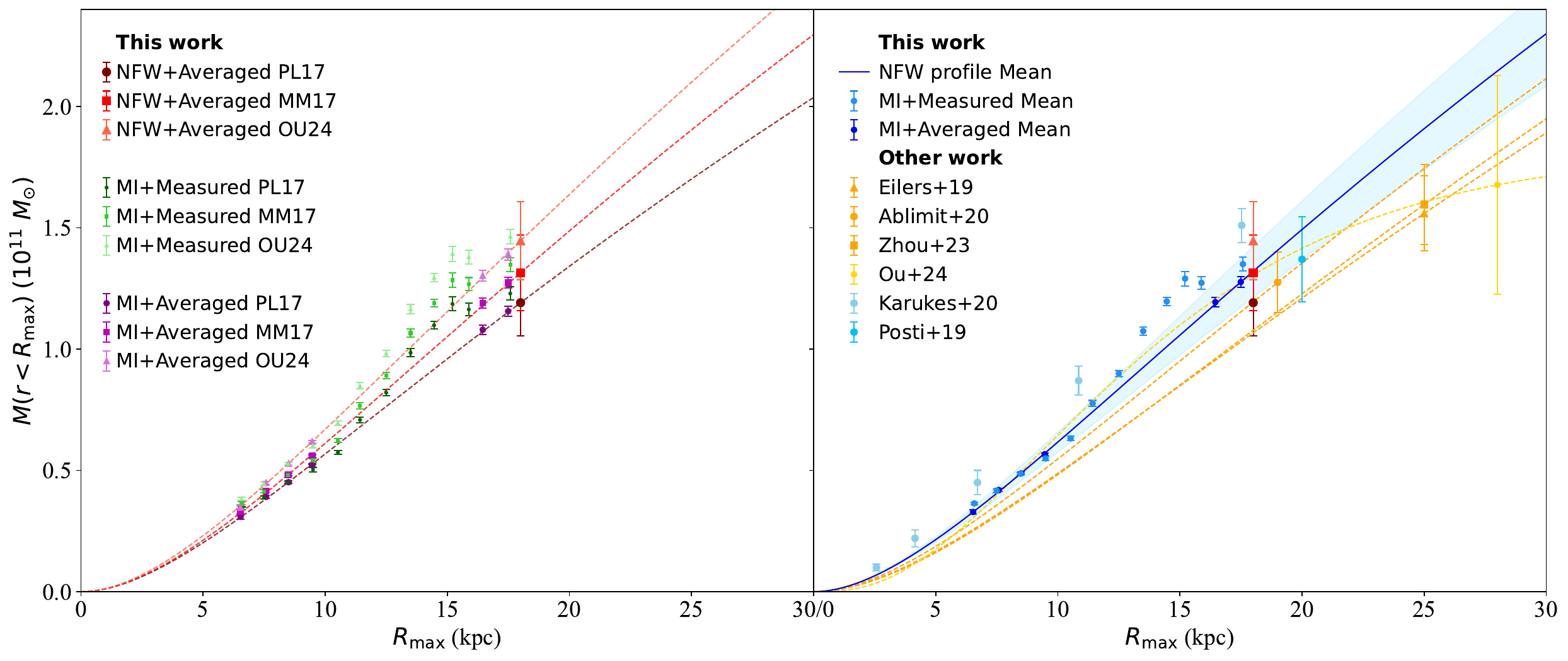}
    \end{center}
    \caption{Left panel: The three red stars mark the enclosed DM halo masses within $18~\mathrm{kpc}$ derived from our best-fit NFW model under three baryonic models adopted from the literature, with the corresponding NFW mass profiles shown as red dashed lines. The green and magenta points represent the DM halo mass profiles derived using the model-independent method (i.e., Equation~7) from our measured RC and the averaged RC, respectively. Right panel: The blue line shows the NFW mass profile derived under the "Mean" baryonic model. The light blue shaded region indicates the $1\sigma$ uncertainty range of the NFW profile, obtained by accounting for the scatter in the RC contributions from the three different literature baryonic models. The dodger blue and blue points correspond to the DM halo mass profiles derived with the model-independent method from our measured RC and the averaged RC, respectively. For comparison, the three orange dashed lines and orange dots indicate the best-fit NFW profiles and the enclosed halo masses within the radial ranges covered by their RCs \protect\citep{EI19, AB20, ZH23}. The gold dashed line and dot represent the best-fit Einasto halo profile and its enclosed mass from \protect\cite{OU24}. The result from \protect\cite{Kar20} is marked with a sky blue dot, while the deep sky blue dot denotes the result from \protect\cite{PH19}, based on the kinematics of globular clusters.}
    \label{Fig5}
\end{figure*}

Compared with previous studies, our measurement of the local DM density is consistent with several RC-based estimates obtained using different tracers \citep{BT12, HU16, EI19, AB20, ZH23}. It is slightly higher than the value derived from the vertical Jeans method \citep{ZA13}. For the enclosed DM mass, the result agrees well with several previous RC-based analyses \citep{EI19, AB20, ZH23}, and it is slightly lower than the estimate of \cite{Kar20}. Our enclosed DM mass measurement is also slightly lower than estimates from studies employing alternative approaches such as the dynamics of tidal streams or globular clusters \citep{KP15, MI18, WT19}, but is consistent with the result obtained by \citet{PH19} using globular cluster kinematics (see Figure 6).

Finally, we acknowledge that, due to the limitations of the current RC determination method, the restricted radial extent of the RC, and potential perturbations from various Galactic structures, the DM profile inferred solely from the RC may be affected by unquantified systematic uncertainties. Strengthening the constraints on the DM profile, particularly in the outer regions, will require incorporating additional probes, such as the escape velocity \citep{PF14} or the kinematics of distant tracers \citep[e.g.][]{WT19, FR20}. Further theoretical developments on DM halo modeling will also be crucial.

\section{Summary}
In this work, we present a new measurement of the Milky Way's RC based on nearly one thousand carefully selected classical Cepheids, a kind of stellar tracer which not only provides precise distance measurements, but also exhibits minimal asymmetric drift. To further reduce its impact, we apply an asymmetric drift correction using the Jeans equation even for these young Cepheids.

Our results confirm that the Galactic RC generally displays a slight downward slope. Specifically, the consistency between our RC and previous measurements based on older tracers demonstrates that the traditional framework adopted in those earlier studies remains reliably applicable within the radial range of $R \lesssim 18\ \mathrm{kpc}$.
 In addition to this overall trend, we observe a local dip-and-bump feature in the RC. This feature has also been identified in other studies tracing the RC \citep[e.g.][]{HU16, MO19, CE19, GC21, ZO24, BE24}, primarily in RCs traced by young tracers such as Cepheids. However, the underlying mechanism responsible for producing this feature remains unclear.

To facilitate subsequent work, we construct an averaged RC by combining measurements from previous efforts with our own to mitigate the effects of systematic uncertainties between different tracers. Additionally, we exclude measurements within the range $10<R<16$~kpc to minimize the influence of the fluctuating feature. Using this averaged RC, we estimate the circular speed at the solar position to be $V_{c,\ \odot}=236.8\pm0.8\ \rm{km s^{-1}}$.

Finally, we constrain the DM halo properties using two approaches. On the one hand, we fit the averaged RC with mass models consisting of several baryonic components and NFW halos. On the other hand, we derive the DM mass profile directly from both our measured RC and the averaged RC in a model-independent manner. From the model fitting, we estimate the enclosed DM halo mass within $18~\mathrm{kpc}$ and the local DM density at the solar position, while the model-independent method yields the DM halo mass profiles. Comparing the derived DM properties under various baryonic models reveals that variations in the assumed baryonic components do not lead to substantial changes in the inferred DM parameters. Overall, both the local DM density and the enclosed DM mass we obtain are in good agreement with previous measurements \citep[e.g.][]{EI19, AB20, ZH23, OU24}.

\section*{Acknowledgements}
We sincerely thank the referee, Dr D. Kawata, for his constructive suggestions that improve the manuscript significantly.

H.W.Z. acknowledges the supported from the National Key R\&D Programme of China (Grant No. 2024YFA1611903). Y.H. acknowledges the supported the National Science Foundation of China (NSFC Grant No. 12422303). 

This work presents results from the European Space Agency’s space mission Gaia (\url{https://www.cosmos.esa.int/gaia}). Gaia data are processed by the Gaia Data Processing and Analysis Consortium (\url{https://www.cosmos.esa.int/web/gaia/dpac/consortium}), which is funded by national institutions, in particular the institutions participating in the Gaia MultiLateral Agreement.

\section*{Data Availability}
The Cepheids data used in this paper are publicly available from the Gaia Archive: \url{https://archives.esac.esa.int/gaia}. 


\bibliography{Bibliography}

@ARTICLE{vA85,
       author = {{van Albada}, T.~S. and {Bahcall}, J.~N. and {Begeman}, K. and {Sancisi}, R.},
        title = "{Distribution of dark matter in the spiral galaxy NGC 3198.}",
      journal = {\apj},
         year = 1985,
        month = aug,
       volume = {295},
        pages = {305-313},
          doi = {10.1086/163375},
       adsurl = {https://ui.adsabs.harvard.edu/abs/1985ApJ...295..305V}
}

@ARTICLE{VD59,
       author = {{Volders}, L.~M.~J.~S.},
        title = "{Neutral hydrogen in M 33 and M 101}",
      journal = {\bain},
         year = 1959,
        month = sep,
       volume = {14},
        pages = {323},
       adsurl = {https://ui.adsabs.harvard.edu/abs/1959BAN....14..323V}
}

@ARTICLE{LA89,
       author = {{Lake}, George},
        title = "{Must the Disk and Halo Dark Matter Be Different?}",
      journal = {\aj},
         year = 1989,
        month = nov,
       volume = {98},
        pages = {1554},
          doi = {10.1086/115238},
       adsurl = {https://ui.adsabs.harvard.edu/abs/1989AJ.....98.1554L}
}

@ARTICLE{DB94,
       author = {{Dubinski}, John},
        title = "{The Effect of Dissipation on the Shapes of Dark Halos}",
      journal = {\apj},
         year = 1994,
        month = aug,
       volume = {431},
        pages = {617},
          doi = {10.1086/174512},
archivePrefix = {arXiv},
       eprint = {astro-ph/9309001},
 primaryClass = {astro-ph},
       adsurl = {https://ui.adsabs.harvard.edu/abs/1994ApJ...431..617D}
}

@ARTICLE{LX12,
       author = {{Lux}, H. and {Read}, J.~I. and {Lake}, G. and {Johnston}, K.~V.},
        title = "{NGC 5466: a unique probe of the Galactic halo shape}",
      journal = {\mnras},
         year = 2012,
        month = jul,
       volume = {424},
       number = {1},
        pages = {L16-L20},
          doi = {10.1111/j.1745-3933.2012.01276.x},
archivePrefix = {arXiv},
       eprint = {1204.5771},
 primaryClass = {astro-ph.GA},
       adsurl = {https://ui.adsabs.harvard.edu/abs/2012MNRAS.424L..16L}
}

@ARTICLE{DE19,
       author = {{Deason}, Alis J. and {Belokurov}, Vasily and {Sanders}, Jason L.},
        title = "{The total stellar halo mass of the Milky Way}",
      journal = {\mnras},
         year = 2019,
        month = dec,
       volume = {490},
       number = {3},
        pages = {3426-3439},
          doi = {10.1093/mnras/stz2793},
archivePrefix = {arXiv},
       eprint = {1908.02763},
 primaryClass = {astro-ph.GA},
       adsurl = {https://ui.adsabs.harvard.edu/abs/2019MNRAS.490.3426D}
}

@ARTICLE{FR70,
       author = {{Freeman}, K.~C.},
        title = "{On the Disks of Spiral and S0 Galaxies}",
      journal = {\apj},
         year = 1970,
        month = jun,
       volume = {160},
        pages = {811},
          doi = {10.1086/150474},
       adsurl = {https://ui.adsabs.harvard.edu/abs/1970ApJ...160..811F}
}

@ARTICLE{RB80,
       author = {{Rubin}, V.~C. and {Ford}, Jr., W.~K. and {Thonnard}, N.},
        title = "{Rotational properties of 21 SC galaxies with a large range of luminosities and radii, from NGC 4605 (R=4kpc) to UGC 2885 (R=122kpc).}",
      journal = {\apj},
         year = 1980,
        month = jun,
       volume = {238},
        pages = {471-487},
          doi = {10.1086/158003},
       adsurl = {https://ui.adsabs.harvard.edu/abs/1980ApJ...238..471R}
}

@ARTICLE{SA10,
       author = {{Salucci}, P. and {Nesti}, F. and {Gentile}, G. and {Frigerio Martins}, C.},
        title = "{The dark matter density at the Sun's location}",
      journal = {\aap},
         year = 2010,
        month = nov,
       volume = {523},
          eid = {A83},
        pages = {A83},
          doi = {10.1051/0004-6361/201014385},
archivePrefix = {arXiv},
       eprint = {1003.3101},
 primaryClass = {astro-ph.GA},
       adsurl = {https://ui.adsabs.harvard.edu/abs/2010A&A...523A..83S}
}

@ARTICLE{WB10,
       author = {{Weber}, M. and {de Boer}, W.},
        title = "{Determination of the local dark matter density in our Galaxy}",
      journal = {\aap},
         year = 2010,
        month = jan,
       volume = {509},
          eid = {A25},
        pages = {A25},
          doi = {10.1051/0004-6361/200913381},
archivePrefix = {arXiv},
       eprint = {0910.4272},
 primaryClass = {astro-ph.CO},
       adsurl = {https://ui.adsabs.harvard.edu/abs/2010A&A...509A..25W}
}

@ARTICLE{BG78,
       author = {{Burton}, W.~B. and {Gordon}, M.~A.},
        title = "{Carbon monoxide in the Galaxy. III. The overall nature of its distribution in the equatorial plane.}",
      journal = {\aap},
         year = 1978,
        month = feb,
       volume = {63},
       number = {1-2},
        pages = {7-27},
       adsurl = {https://ui.adsabs.harvard.edu/abs/1978A&A....63....7B}
}

@ARTICLE{GU79,
       author = {{Gunn}, J.~E. and {Knapp}, G.~R. and {Tremaine}, S.~D.},
        title = "{The global properties of the Galaxy. II. The galactic rotation parameters from 21-cm H I observations.}",
      journal = {\aj},
         year = 1979,
        month = aug,
       volume = {84},
        pages = {1181-1188},
          doi = {10.1086/112525},
       adsurl = {https://ui.adsabs.harvard.edu/abs/1979AJ.....84.1181G}
}

@ARTICLE{FI89,
       author = {{Fich}, Michel and {Blitz}, Leo and {Stark}, Antony A.},
        title = "{The Rotation Curve of the Milky Way to 2R 0}",
      journal = {\apj},
         year = 1989,
        month = jul,
       volume = {342},
        pages = {272},
          doi = {10.1086/167591},
       adsurl = {https://ui.adsabs.harvard.edu/abs/1989ApJ...342..272F}
}

@ARTICLE{LE08,
       author = {{Levine}, E.~S. and {Heiles}, Carl and {Blitz}, Leo},
        title = "{The Milky Way Rotation Curve and Its Vertical Derivatives: Inside the Solar Circle}",
      journal = {\apj},
         year = 2008,
        month = jun,
       volume = {679},
       number = {2},
        pages = {1288-1298},
          doi = {10.1086/587444},
archivePrefix = {arXiv},
       eprint = {0802.2714},
 primaryClass = {astro-ph},
       adsurl = {https://ui.adsabs.harvard.edu/abs/2008ApJ...679.1288L}
}

@ARTICLE{SO09,
       author = {{Sofue}, Yoshiaki and {Honma}, Mareki and {Omodaka}, Toshihiro},
        title = "{Unified Rotation Curve of the Galaxy -- Decomposition into de Vaucouleurs Bulge, Disk, Dark Halo, and the 9-kpc Rotation Dip --}",
      journal = {\pasj},
         year = 2009,
        month = feb,
       volume = {61},
        pages = {227},
          doi = {10.1093/pasj/61.2.227},
archivePrefix = {arXiv},
       eprint = {0811.0859},
 primaryClass = {astro-ph},
       adsurl = {https://ui.adsabs.harvard.edu/abs/2009PASJ...61..227S}
}

@ARTICLE{SO091,
       author = {{Sofue}, Yoshiaki},
        title = "{Pseudo Rotation Curve Connecting the Galaxy, Dark Halo, and Local Group}",
      journal = {\pasj},
         year = 2009,
        month = feb,
       volume = {61},
        pages = {153},
          doi = {10.1093/pasj/61.2.153},
archivePrefix = {arXiv},
       eprint = {0811.0860},
 primaryClass = {astro-ph},
       adsurl = {https://ui.adsabs.harvard.edu/abs/2009PASJ...61..153S}
}

@ARTICLE{PT97,
       author = {{Pont}, F. and {Queloz}, D. and {Bratschi}, P. and {Mayor}, M.},
        title = "{Rotation of the outer disc from classical cepheids.}",
      journal = {\aap},
         year = 1997,
        month = feb,
       volume = {318},
        pages = {416-428},
       adsurl = {https://ui.adsabs.harvard.edu/abs/1997A&A...318..416P}
}

@ARTICLE{MO19,
       author = {{Mr{\'o}z}, Przemek and {Udalski}, Andrzej and {Skowron}, Dorota M. and {Skowron}, Jan and {Soszy{\'n}ski}, Igor and {Pietrukowicz}, Pawe{\l} and {Szyma{\'n}ski}, Micha{\l} K. and {Poleski}, Rados{\l}aw and {Koz{\l}owski}, Szymon and {Ulaczyk}, Krzysztof},
        title = "{Rotation Curve of the Milky Way from Classical Cepheids}",
      journal = {\apjl},
         year = 2019,
        month = jan,
       volume = {870},
       number = {1},
          eid = {L10},
        pages = {L10},
          doi = {10.3847/2041-8213/aaf73f},
archivePrefix = {arXiv},
       eprint = {1810.02131},
 primaryClass = {astro-ph.GA},
       adsurl = {https://ui.adsabs.harvard.edu/abs/2019ApJ...870L..10M}
}

@ARTICLE{KW19,
       author = {{Kawata}, Daisuke and {Bovy}, Jo and {Matsunaga}, Noriyuki and {Baba}, Junichi},
        title = "{Galactic rotation from Cepheids with Gaia DR2 and effects of non-axisymmetry}",
      journal = {\mnras},
         year = 2019,
        month = jan,
       volume = {482},
       number = {1},
        pages = {40-51},
          doi = {10.1093/mnras/sty2623},
archivePrefix = {arXiv},
       eprint = {1803.05927},
 primaryClass = {astro-ph.GA},
       adsurl = {https://ui.adsabs.harvard.edu/abs/2019MNRAS.482...40K}
}

@ARTICLE{AB20,
       author = {{Ablimit}, Iminhaji and {Zhao}, Gang and {Flynn}, Chris and {Bird}, Sarah A.},
        title = "{The Rotation Curve, Mass Distribution, and Dark Matter Content of the Milky Way from Classical Cepheids}",
      journal = {\apjl},
         year = 2020,
        month = may,
       volume = {895},
       number = {1},
          eid = {L12},
        pages = {L12},
          doi = {10.3847/2041-8213/ab8d45},
archivePrefix = {arXiv},
       eprint = {2004.13768},
 primaryClass = {astro-ph.GA},
       adsurl = {https://ui.adsabs.harvard.edu/abs/2020ApJ...895L..12A}
}

@ARTICLE{BV12b,
       author = {{Bovy}, Jo and {Allende Prieto}, Carlos and {Beers}, Timothy C. and {Bizyaev}, Dmitry and {da Costa}, Luiz N. and {Cunha}, Katia and {Ebelke}, Garrett L. and {Eisenstein}, Daniel J. and {Frinchaboy}, Peter M. and {Garc{\'\i}a P{\'e}rez}, Ana Elia and {Girardi}, L{\'e}o and {Hearty}, Fred R. and {Hogg}, David W. and {Holtzman}, Jon and {Maia}, Marcio A.~G. and {Majewski}, Steven R. and {Malanushenko}, Elena and {Malanushenko}, Viktor and {M{\'e}sz{\'a}ros}, Szabolcs and {Nidever}, David L. and {O'Connell}, Robert W. and {O'Donnell}, Christine and {Oravetz}, Audrey and {Pan}, Kaike and {Rocha-Pinto}, Helio J. and {Schiavon}, Ricardo P. and {Schneider}, Donald P. and {Schultheis}, Mathias and {Skrutskie}, Michael and {Smith}, Verne V. and {Weinberg}, David H. and {Wilson}, John C. and {Zasowski}, Gail},
        title = "{The Milky Way's Circular-velocity Curve between 4 and 14 kpc from APOGEE data}",
      journal = {\apj},
         year = 2012,
        month = nov,
       volume = {759},
       number = {2},
          eid = {131},
        pages = {131},
          doi = {10.1088/0004-637X/759/2/131},
archivePrefix = {arXiv},
       eprint = {1209.0759},
 primaryClass = {astro-ph.GA},
       adsurl = {https://ui.adsabs.harvard.edu/abs/2012ApJ...759..131B}
}

@ARTICLE{HU16,
       author = {{Huang}, Y. and {Liu}, X. -W. and {Yuan}, H. -B. and {Xiang}, M. -S. and {Zhang}, H. -W. and {Chen}, B. -Q. and {Ren}, J. -J. and {Wang}, C. and {Zhang}, Y. and {Hou}, Y. -H. and {Wang}, Y. -F. and {Cao}, Z. -H.},
        title = "{The Milky Way's rotation curve out to 100 kpc and its constraint on the Galactic mass distribution}",
      journal = {\mnras},
         year = 2016,
        month = dec,
       volume = {463},
       number = {3},
        pages = {2623-2639},
          doi = {10.1093/mnras/stw2096},
archivePrefix = {arXiv},
       eprint = {1604.01216},
 primaryClass = {astro-ph.GA},
       adsurl = {https://ui.adsabs.harvard.edu/abs/2016MNRAS.463.2623H}
}

@ARTICLE{BB15,
       author = {{Bobylev}, V.~V. and {Bajkova}, A.~T.},
        title = "{Determination of the galactic rotation curve from OB stars}",
      journal = {Astronomy Letters},
         year = 2015,
        month = sep,
       volume = {41},
       number = {9},
        pages = {473-488},
          doi = {10.1134/S1063773715080010},
archivePrefix = {arXiv},
       eprint = {1507.08049},
 primaryClass = {astro-ph.GA},
       adsurl = {https://ui.adsabs.harvard.edu/abs/2015AstL...41..473B}
}

@ARTICLE{BT13,
       author = {{Battinelli}, P. and {Demers*}, S. and {Rossi**}, C. and {Gigoyan}, K.~S.},
        title = "{Extension of the C Star Rotation Curve of the Milky Way to 24 kpc}",
      journal = {Astrophysics},
         year = 2013,
        month = mar,
       volume = {56},
       number = {1},
        pages = {68-75},
          doi = {10.1007/s10511-013-9268-7},
archivePrefix = {arXiv},
       eprint = {1212.1116},
 primaryClass = {astro-ph.SR},
       adsurl = {https://ui.adsabs.harvard.edu/abs/2013Ap.....56...68B}
}

@ARTICLE{BB93,
       author = {{Brand}, J. and {Blitz}, L.},
        title = "{The velocity field of the outer galaxy.}",
      journal = {\aap},
         year = 1993,
        month = aug,
       volume = {275},
        pages = {67-90},
       adsurl = {https://ui.adsabs.harvard.edu/abs/1993A&A...275...67B}
}

@ARTICLE{RE14,
       author = {{Reid}, M.~J. and {Menten}, K.~M. and {Brunthaler}, A. and {Zheng}, X.~W. and {Dame}, T.~M. and {Xu}, Y. and {Wu}, Y. and {Zhang}, B. and {Sanna}, A. and {Sato}, M. and {Hachisuka}, K. and {Choi}, Y.~K. and {Immer}, K. and {Moscadelli}, L. and {Rygl}, K.~L.~J. and {Bartkiewicz}, A.},
        title = "{Trigonometric Parallaxes of High Mass Star Forming Regions: The Structure and Kinematics of the Milky Way}",
      journal = {\apj},
         year = 2014,
        month = mar,
       volume = {783},
       number = {2},
          eid = {130},
        pages = {130},
          doi = {10.1088/0004-637X/783/2/130},
archivePrefix = {arXiv},
       eprint = {1401.5377},
 primaryClass = {astro-ph.GA},
       adsurl = {https://ui.adsabs.harvard.edu/abs/2014ApJ...783..130R}
}

@ARTICLE{XU08,
       author = {{Xue}, X.~X. and {Rix}, H.~W. and {Zhao}, G. and {Re Fiorentin}, P. and {Naab}, T. and {Steinmetz}, M. and {van den Bosch}, F.~C. and {Beers}, T.~C. and {Lee}, Y.~S. and {Bell}, E.~F. and {Rockosi}, C. and {Yanny}, B. and {Newberg}, H. and {Wilhelm}, R. and {Kang}, X. and {Smith}, M.~C. and {Schneider}, D.~P.},
        title = "{The Milky Way's Circular Velocity Curve to 60 kpc and an Estimate of the Dark Matter Halo Mass from the Kinematics of \raisebox{-0.5ex}\textasciitilde2400 SDSS Blue Horizontal-Branch Stars}",
      journal = {\apj},
         year = 2008,
        month = sep,
       volume = {684},
       number = {2},
        pages = {1143-1158},
          doi = {10.1086/589500},
archivePrefix = {arXiv},
       eprint = {0801.1232},
 primaryClass = {astro-ph},
       adsurl = {https://ui.adsabs.harvard.edu/abs/2008ApJ...684.1143X}
}

@ARTICLE{KF12,
       author = {{Kafle}, Prajwal R. and {Sharma}, Sanjib and {Lewis}, Geraint F. and {Bland-Hawthorn}, Joss},
        title = "{Kinematics of the Stellar Halo and the Mass Distribution of the Milky Way Using Blue Horizontal Branch Stars}",
      journal = {\apj},
         year = 2012,
        month = dec,
       volume = {761},
       number = {2},
          eid = {98},
        pages = {98},
          doi = {10.1088/0004-637X/761/2/98},
archivePrefix = {arXiv},
       eprint = {1210.7527},
 primaryClass = {astro-ph.GA},
       adsurl = {https://ui.adsabs.harvard.edu/abs/2012ApJ...761...98K}
}

@ARTICLE{HR87,
       author = {{Hron}, J.},
        title = "{Kinematics of young open clusters and the rotation curve of our Galaxy.}",
      journal = {\aap},
         year = 1987,
        month = apr,
       volume = {176},
        pages = {34-52},
       adsurl = {https://ui.adsabs.harvard.edu/abs/1987A&A...176...34H}
}

@ARTICLE{SF09,
       author = {{Spitler}, L.~R. and {Forbes}, D.~A.},
        title = "{A new method for estimating dark matter halo masses using globular cluster systems}",
      journal = {\mnras},
         year = 2009,
        month = jan,
       volume = {392},
       number = {1},
        pages = {L1-L5},
          doi = {10.1111/j.1745-3933.2008.00567.x},
archivePrefix = {arXiv},
       eprint = {0809.5057},
 primaryClass = {astro-ph},
       adsurl = {https://ui.adsabs.harvard.edu/abs/2009MNRAS.392L...1S}
}

@ARTICLE{BG05,
       author = {{Battaglia}, Giuseppina and {Helmi}, Amina and {Morrison}, Heather and {Harding}, Paul and {Olszewski}, Edward W. and {Mateo}, Mario and {Freeman}, Kenneth C. and {Norris}, John and {Shectman}, Stephen A.},
        title = "{The radial velocity dispersion profile of the Galactic halo: constraining the density profile of the dark halo of the Milky Way}",
      journal = {\mnras},
         year = 2005,
        month = dec,
       volume = {364},
       number = {2},
        pages = {433-442},
          doi = {10.1111/j.1365-2966.2005.09367.x},
archivePrefix = {arXiv},
       eprint = {astro-ph/0506102},
 primaryClass = {astro-ph},
       adsurl = {https://ui.adsabs.harvard.edu/abs/2005MNRAS.364..433B}
}

@ARTICLE{GC16,
       author = {{Gaia Collaboration} and {Brown}, A.~G.~A. and {Vallenari}, A. and {Prusti}, T. and {de Bruijne}, J.~H.~J. and {Mignard}, F. and {Drimmel}, R. and {Babusiaux}, C. and {Bailer-Jones}, C.~A.~L. and {Bastian}, U. and {Biermann}, M. and {Evans}, D.~W. and {Eyer}, L. and {Jansen}, F. and {Jordi}, C. and {Katz}, D. and {Klioner}, S.~A. and {Lammers}, U. and {Lindegren}, L. and {Luri}, X. and {O'Mullane}, W. and {Panem}, C. and {Pourbaix}, D. and {Randich}, S. and {Sartoretti}, P. and {Siddiqui}, H.~I. and {Soubiran}, C. and {Valette}, V. and {van Leeuwen}, F. and {Walton}, N.~A. and {Aerts}, C. and {Arenou}, F. and {Cropper}, M. and {H{\o}g}, E. and {Lattanzi}, M.~G. and {Grebel}, E.~K. and {Holland}, A.~D. and {Huc}, C. and {Passot}, X. and {Perryman}, M. and {Bramante}, L. and {Cacciari}, C. and {Casta{\~n}eda}, J. and {Chaoul}, L. and {Cheek}, N. and {De Angeli}, F. and {Fabricius}, C. and {Guerra}, R. and {Hern{\'a}ndez}, J. and {Jean-Antoine-Piccolo}, A. and {Masana}, E. and {Messineo}, R. and {Mowlavi}, N. and {Nienartowicz}, K. and {Ord{\'o}{\~n}ez-Blanco}, D. and {Panuzzo}, P. and {Portell}, J. and {Richards}, P.~J. and {Riello}, M. and {Seabroke}, G.~M. and {Tanga}, P. and {Th{\'e}venin}, F. and {Torra}, J. and {Els}, S.~G. and {Gracia-Abril}, G. and {Comoretto}, G. and {Garcia-Reinaldos}, M. and {Lock}, T. and {Mercier}, E. and {Altmann}, M. and {Andrae}, R. and {Astraatmadja}, T.~L. and {Bellas-Velidis}, I. and {Benson}, K. and {Berthier}, J. and {Blomme}, R. and {Busso}, G. and {Carry}, B. and {Cellino}, A. and {Clementini}, G. and {Cowell}, S. and {Creevey}, O. and {Cuypers}, J. and {Davidson}, M. and {De Ridder}, J. and {de Torres}, A. and {Delchambre}, L. and {Dell'Oro}, A. and {Ducourant}, C. and {Fr{\'e}mat}, Y. and {Garc{\'\i}a-Torres}, M. and {Gosset}, E. and {Halbwachs}, J. -L. and {Hambly}, N.~C. and {Harrison}, D.~L. and {Hauser}, M. and {Hestroffer}, D. and {Hodgkin}, S.~T. and {Huckle}, H.~E. and {Hutton}, A. and {Jasniewicz}, G. and {Jordan}, S. and {Kontizas}, M. and {Korn}, A.~J. and {Lanzafame}, A.~C. and {Manteiga}, M. and {Moitinho}, A. and {Muinonen}, K. and {Osinde}, J. and {Pancino}, E. and {Pauwels}, T. and {Petit}, J. -M. and {Recio-Blanco}, A. and {Robin}, A.~C. and {Sarro}, L.~M. and {Siopis}, C. and {Smith}, M. and {Smith}, K.~W. and {Sozzetti}, A. and {Thuillot}, W. and {van Reeven}, W. and {Viala}, Y. and {Abbas}, U. and {Abreu Aramburu}, A. and {Accart}, S. and {Aguado}, J.~J. and {Allan}, P.~M. and {Allasia}, W. and {Altavilla}, G. and {{\'A}lvarez}, M.~A. and {Alves}, J. and {Anderson}, R.~I. and {Andrei}, A.~H. and {Anglada Varela}, E. and {Antiche}, E. and {Antoja}, T. and {Ant{\'o}n}, S. and {Arcay}, B. and {Bach}, N. and {Baker}, S.~G. and {Balaguer-N{\'u}{\~n}ez}, L. and {Barache}, C. and {Barata}, C. and {Barbier}, A. and {Barblan}, F. and {Barrado y Navascu{\'e}s}, D. and {Barros}, M. and {Barstow}, M.~A. and {Becciani}, U. and {Bellazzini}, M. and {Bello Garc{\'\i}a}, A. and {Belokurov}, V. and {Bendjoya}, P. and {Berihuete}, A. and {Bianchi}, L. and {Bienaym{\'e}}, O. and {Billebaud}, F. and {Blagorodnova}, N. and {Blanco-Cuaresma}, S. and {Boch}, T. and {Bombrun}, A. and {Borrachero}, R. and {Bouquillon}, S. and {Bourda}, G. and {Bouy}, H. and {Bragaglia}, A. and {Breddels}, M.~A. and {Brouillet}, N. and {Br{\"u}semeister}, T. and {Bucciarelli}, B. and {Burgess}, P. and {Burgon}, R. and {Burlacu}, A. and {Busonero}, D. and {Buzzi}, R. and {Caffau}, E. and {Cambras}, J. and {Campbell}, H. and {Cancelliere}, R. and {Cantat-Gaudin}, T. and {Carlucci}, T. and {Carrasco}, J.~M. and {Castellani}, M. and {Charlot}, P. and {Charnas}, J. and {Chiavassa}, A. and {Clotet}, M. and {Cocozza}, G. and {Collins}, R.~S. and {Costigan}, G. and {Crifo}, F. and {Cross}, N.~J.~G. and {Crosta}, M. and {Crowley}, C. and {Dafonte}, C. and {Damerdji}, Y. and {Dapergolas}, A. and {David}, P. and {David}, M. and {De Cat}, P.},
        title = "{Gaia Data Release 1. Summary of the astrometric, photometric, and survey properties}",
      journal = {\aap},
         year = 2016,
        month = nov,
       volume = {595},
          eid = {A2},
        pages = {A2},
          doi = {10.1051/0004-6361/201629512},
archivePrefix = {arXiv},
       eprint = {1609.04172},
 primaryClass = {astro-ph.IM},
       adsurl = {https://ui.adsabs.harvard.edu/abs/2016A&A...595A...2G}
}

@ARTICLE{GC18,
       author = {{Gaia Collaboration} and {Brown}, A.~G.~A. and {Vallenari}, A. and {Prusti}, T. and {de Bruijne}, J.~H.~J. and {Babusiaux}, C. and {Bailer-Jones}, C.~A.~L. and {Biermann}, M. and {Evans}, D.~W. and {Eyer}, L. and {Jansen}, F. and {Jordi}, C. and {Klioner}, S.~A. and {Lammers}, U. and {Lindegren}, L. and {Luri}, X. and {Mignard}, F. and {Panem}, C. and {Pourbaix}, D. and {Randich}, S. and {Sartoretti}, P. and {Siddiqui}, H.~I. and {Soubiran}, C. and {van Leeuwen}, F. and {Walton}, N.~A. and {Arenou}, F. and {Bastian}, U. and {Cropper}, M. and {Drimmel}, R. and {Katz}, D. and {Lattanzi}, M.~G. and {Bakker}, J. and {Cacciari}, C. and {Casta{\~n}eda}, J. and {Chaoul}, L. and {Cheek}, N. and {De Angeli}, F. and {Fabricius}, C. and {Guerra}, R. and {Holl}, B. and {Masana}, E. and {Messineo}, R. and {Mowlavi}, N. and {Nienartowicz}, K. and {Panuzzo}, P. and {Portell}, J. and {Riello}, M. and {Seabroke}, G.~M. and {Tanga}, P. and {Th{\'e}venin}, F. and {Gracia-Abril}, G. and {Comoretto}, G. and {Garcia-Reinaldos}, M. and {Teyssier}, D. and {Altmann}, M. and {Andrae}, R. and {Audard}, M. and {Bellas-Velidis}, I. and {Benson}, K. and {Berthier}, J. and {Blomme}, R. and {Burgess}, P. and {Busso}, G. and {Carry}, B. and {Cellino}, A. and {Clementini}, G. and {Clotet}, M. and {Creevey}, O. and {Davidson}, M. and {De Ridder}, J. and {Delchambre}, L. and {Dell'Oro}, A. and {Ducourant}, C. and {Fern{\'a}ndez-Hern{\'a}ndez}, J. and {Fouesneau}, M. and {Fr{\'e}mat}, Y. and {Galluccio}, L. and {Garc{\'\i}a-Torres}, M. and {Gonz{\'a}lez-N{\'u}{\~n}ez}, J. and {Gonz{\'a}lez-Vidal}, J.~J. and {Gosset}, E. and {Guy}, L.~P. and {Halbwachs}, J. -L. and {Hambly}, N.~C. and {Harrison}, D.~L. and {Hern{\'a}ndez}, J. and {Hestroffer}, D. and {Hodgkin}, S.~T. and {Hutton}, A. and {Jasniewicz}, G. and {Jean-Antoine-Piccolo}, A. and {Jordan}, S. and {Korn}, A.~J. and {Krone-Martins}, A. and {Lanzafame}, A.~C. and {Lebzelter}, T. and {L{\"o}ffler}, W. and {Manteiga}, M. and {Marrese}, P.~M. and {Mart{\'\i}n-Fleitas}, J.~M. and {Moitinho}, A. and {Mora}, A. and {Muinonen}, K. and {Osinde}, J. and {Pancino}, E. and {Pauwels}, T. and {Petit}, J. -M. and {Recio-Blanco}, A. and {Richards}, P.~J. and {Rimoldini}, L. and {Robin}, A.~C. and {Sarro}, L.~M. and {Siopis}, C. and {Smith}, M. and {Sozzetti}, A. and {S{\"u}veges}, M. and {Torra}, J. and {van Reeven}, W. and {Abbas}, U. and {Abreu Aramburu}, A. and {Accart}, S. and {Aerts}, C. and {Altavilla}, G. and {{\'A}lvarez}, M.~A. and {Alvarez}, R. and {Alves}, J. and {Anderson}, R.~I. and {Andrei}, A.~H. and {Anglada Varela}, E. and {Antiche}, E. and {Antoja}, T. and {Arcay}, B. and {Astraatmadja}, T.~L. and {Bach}, N. and {Baker}, S.~G. and {Balaguer-N{\'u}{\~n}ez}, L. and {Balm}, P. and {Barache}, C. and {Barata}, C. and {Barbato}, D. and {Barblan}, F. and {Barklem}, P.~S. and {Barrado}, D. and {Barros}, M. and {Barstow}, M.~A. and {Bartholom{\'e} Mu{\~n}oz}, S. and {Bassilana}, J. -L. and {Becciani}, U. and {Bellazzini}, M. and {Berihuete}, A. and {Bertone}, S. and {Bianchi}, L. and {Bienaym{\'e}}, O. and {Blanco-Cuaresma}, S. and {Boch}, T. and {Boeche}, C. and {Bombrun}, A. and {Borrachero}, R. and {Bossini}, D. and {Bouquillon}, S. and {Bourda}, G. and {Bragaglia}, A. and {Bramante}, L. and {Breddels}, M.~A. and {Bressan}, A. and {Brouillet}, N. and {Br{\"u}semeister}, T. and {Brugaletta}, E. and {Bucciarelli}, B. and {Burlacu}, A. and {Busonero}, D. and {Butkevich}, A.~G. and {Buzzi}, R. and {Caffau}, E. and {Cancelliere}, R. and {Cannizzaro}, G. and {Cantat-Gaudin}, T. and {Carballo}, R. and {Carlucci}, T. and {Carrasco}, J.~M. and {Casamiquela}, L. and {Castellani}, M. and {Castro-Ginard}, A. and {Charlot}, P. and {Chemin}, L. and {Chiavassa}, A. and {Cocozza}, G. and {Costigan}, G. and {Cowell}, S. and {Crifo}, F. and {Crosta}, M. and {Crowley}, C. and {Cuypers}, J. and {Dafonte}, C. and {Damerdji}, Y. and {Dapergolas}, A. and {David}, P. and {David}, M. and {de Laverny}, P. and {De Luise}, F.},
        title = "{Gaia Data Release 2. Summary of the contents and survey properties}",
      journal = {\aap},
         year = 2018,
        month = aug,
       volume = {616},
          eid = {A1},
        pages = {A1},
          doi = {10.1051/0004-6361/201833051},
archivePrefix = {arXiv},
       eprint = {1804.09365},
 primaryClass = {astro-ph.GA},
       adsurl = {https://ui.adsabs.harvard.edu/abs/2018A&A...616A...1G}
}

@ARTICLE{HG19,
       author = {{Hogg}, David W. and {Eilers}, Anna-Christina and {Rix}, Hans-Walter},
        title = "{Spectrophotometric Parallaxes with Linear Models: Accurate Distances for Luminous Red-giant Stars}",
      journal = {\aj},
         year = 2019,
        month = oct,
       volume = {158},
       number = {4},
          eid = {147},
        pages = {147},
          doi = {10.3847/1538-3881/ab398c},
archivePrefix = {arXiv},
       eprint = {1810.09468},
 primaryClass = {astro-ph.GA},
       adsurl = {https://ui.adsabs.harvard.edu/abs/2019AJ....158..147H}
}

@ARTICLE{ZH23,
       author = {{Zhou}, Yuan and {Li}, Xinyi and {Huang}, Yang and {Zhang}, Huawei},
        title = "{The Circular Velocity Curve of the Milky Way from 5-25 kpc Using Luminous Red Giant Branch Stars}",
      journal = {\apj},
         year = 2023,
        month = apr,
       volume = {946},
       number = {2},
          eid = {73},
        pages = {73},
          doi = {10.3847/1538-4357/acadd9},
archivePrefix = {arXiv},
       eprint = {2212.10393},
 primaryClass = {astro-ph.GA},
       adsurl = {https://ui.adsabs.harvard.edu/abs/2023ApJ...946...73Z}
}

@ARTICLE{MJ17,
       author = {{Majewski}, Steven R. and {Schiavon}, Ricardo P. and {Frinchaboy}, Peter M. and {Allende Prieto}, Carlos and {Barkhouser}, Robert and {Bizyaev}, Dmitry and {Blank}, Basil and {Brunner}, Sophia and {Burton}, Adam and {Carrera}, Ricardo and {Chojnowski}, S. Drew and {Cunha}, K{\'a}tia and {Epstein}, Courtney and {Fitzgerald}, Greg and {Garc{\'\i}a P{\'e}rez}, Ana E. and {Hearty}, Fred R. and {Henderson}, Chuck and {Holtzman}, Jon A. and {Johnson}, Jennifer A. and {Lam}, Charles R. and {Lawler}, James E. and {Maseman}, Paul and {M{\'e}sz{\'a}ros}, Szabolcs and {Nelson}, Matthew and {Nguyen}, Duy Coung and {Nidever}, David L. and {Pinsonneault}, Marc and {Shetrone}, Matthew and {Smee}, Stephen and {Smith}, Verne V. and {Stolberg}, Todd and {Skrutskie}, Michael F. and {Walker}, Eric and {Wilson}, John C. and {Zasowski}, Gail and {Anders}, Friedrich and {Basu}, Sarbani and {Beland}, Stephane and {Blanton}, Michael R. and {Bovy}, Jo and {Brownstein}, Joel R. and {Carlberg}, Joleen and {Chaplin}, William and {Chiappini}, Cristina and {Eisenstein}, Daniel J. and {Elsworth}, Yvonne and {Feuillet}, Diane and {Fleming}, Scott W. and {Galbraith-Frew}, Jessica and {Garc{\'\i}a}, Rafael A. and {Garc{\'\i}a-Hern{\'a}ndez}, D. An{\'\i}bal and {Gillespie}, Bruce A. and {Girardi}, L{\'e}o and {Gunn}, James E. and {Hasselquist}, Sten and {Hayden}, Michael R. and {Hekker}, Saskia and {Ivans}, Inese and {Kinemuchi}, Karen and {Klaene}, Mark and {Mahadevan}, Suvrath and {Mathur}, Savita and {Mosser}, Beno{\^\i}t and {Muna}, Demitri and {Munn}, Jeffrey A. and {Nichol}, Robert C. and {O'Connell}, Robert W. and {Parejko}, John K. and {Robin}, A.~C. and {Rocha-Pinto}, Helio and {Schultheis}, Matthias and {Serenelli}, Aldo M. and {Shane}, Neville and {Silva Aguirre}, Victor and {Sobeck}, Jennifer S. and {Thompson}, Benjamin and {Troup}, Nicholas W. and {Weinberg}, David H. and {Zamora}, Olga},
        title = "{The Apache Point Observatory Galactic Evolution Experiment (APOGEE)}",
      journal = {\aj},
         year = 2017,
        month = sep,
       volume = {154},
       number = {3},
          eid = {94},
        pages = {94},
          doi = {10.3847/1538-3881/aa784d},
archivePrefix = {arXiv},
       eprint = {1509.05420},
 primaryClass = {astro-ph.IM},
       adsurl = {https://ui.adsabs.harvard.edu/abs/2017AJ....154...94M}
}

@ARTICLE{LU15,
       author = {{Luo}, A. -Li and {Zhao}, Yong-Heng and {Zhao}, Gang and {Deng}, Li-Cai and {Liu}, Xiao-Wei and {Jing}, Yi-Peng and {Wang}, Gang and {Zhang}, Hao-Tong and {Shi}, Jian-Rong and {Cui}, Xiang-Qun and {Chu}, Yao-Quan and {Li}, Guo-Ping and {Bai}, Zhong-Rui and {Wu}, Yue and {Cai}, Yan and {Cao}, Shu-Yun and {Cao}, Zi-Huang and {Carlin}, Jeffrey L. and {Chen}, Hai-Yuan and {Chen}, Jian-Jun and {Chen}, Kun-Xin and {Chen}, Li and {Chen}, Xue-Lei and {Chen}, Xiao-Yan and {Chen}, Ying and {Christlieb}, Norbert and {Chu}, Jia-Ru and {Cui}, Chen-Zhou and {Dong}, Yi-Qiao and {Du}, Bing and {Fan}, Dong-Wei and {Feng}, Lei and {Fu}, Jian-Ning and {Gao}, Peng and {Gong}, Xue-Fei and {Gu}, Bo-Zhong and {Guo}, Yan-Xin and {Han}, Zhan-Wen and {He}, Bo-Liang and {Hou}, Jin-Liang and {Hou}, Yong-Hui and {Hou}, Wen and {Hu}, Hong-Zhuan and {Hu}, Ning-Sheng and {Hu}, Zhong-Wen and {Huo}, Zhi-Ying and {Jia}, Lei and {Jiang}, Fang-Hua and {Jiang}, Xiang and {Jiang}, Zhi-Bo and {Jin}, Ge and {Kong}, Xiao and {Kong}, Xu and {Lei}, Ya-Juan and {Li}, Ai-Hua and {Li}, Chang-Hua and {Li}, Guang-Wei and {Li}, Hai-Ning and {Li}, Jian and {Li}, Qi and {Li}, Shuang and {Li}, Sha-Sha and {Li}, Xin-Nan and {Li}, Yan and {Li}, Yin-Bi and {Li}, Ye-Ping and {Liang}, Yuan and {Lin}, Chien-Cheng and {Liu}, Chao and {Liu}, Gen-Rong and {Liu}, Guan-Qun and {Liu}, Zhi-Gang and {Lu}, Wen-Zhi and {Luo}, Yu and {Mao}, Yin-Dun and {Newberg}, Heidi and {Ni}, Ji-Jun and {Qi}, Zhao-Xiang and {Qi}, Yong-Jun and {Shen}, Shi-Yin and {Shi}, Huo-Ming and {Song}, Jing and {Song}, Yi-Han and {Su}, Ding-Qiang and {Su}, Hong-Jun and {Tang}, Zheng-Hong and {Tao}, Qing-Sheng and {Tian}, Yuan and {Wang}, Dan and {Wang}, Da-Qi and {Wang}, Feng-Fei and {Wang}, Guo-Min and {Wang}, Hai and {Wang}, Hong-Chi and {Wang}, Jian and {Wang}, Jia-Ning and {Wang}, Jian-Ling and {Wang}, Jian-Ping and {Wang}, Jun-Xian and {Wang}, Lei and {Wang}, Meng-Xin and {Wang}, Shou-Guan and {Wang}, Shu-Qing and {Wang}, Xia and {Wang}, Ya-Nan and {Wang}, You and {Wang}, Yue-Fei and {Wang}, You-Fen and {Wei}, Peng and {Wei}, Ming-Zhi and {Wu}, Hong and {Wu}, Ke-Fei and {Wu}, Xue-Bing and {Wu}, Yu-Zhong and {Xing}, Xiao-Zheng and {Xu}, Ling-Zhe and {Xu}, Xin-Qi and {Xu}, Yan and {Yan}, Tai-Sheng and {Yang}, De-Hua and {Yang}, Hai-Feng and {Yang}, Hui-Qin and {Yang}, Ming and {Yao}, Zheng-Qiu and {Yu}, Yong and {Yuan}, Hui and {Yuan}, Hai-Bo and {Yuan}, Hai-Long and {Yuan}, Wei-Min and {Zhai}, Chao and {Zhang}, En-Peng and {Zhang}, Hua-Wei and {Zhang}, Jian-Nan and {Zhang}, Li-Pin and {Zhang}, Wei and {Zhang}, Yong and {Zhang}, Yan-Xia and {Zhang}, Zheng-Chao and {Zhao}, Ming and {Zhou}, Fang and {Zhou}, Xu and {Zhu}, Jie and {Zhu}, Yong-Tian and {Zou}, Si-Cheng and {Zuo}, Fang},
        title = "{The first data release (DR1) of the LAMOST regular survey}",
      journal = {Research in Astronomy and Astrophysics},
         year = 2015,
        month = aug,
       volume = {15},
       number = {8},
          eid = {1095},
        pages = {1095},
          doi = {10.1088/1674-4527/15/8/002},
archivePrefix = {arXiv},
       eprint = {1505.01570},
 primaryClass = {astro-ph.GA},
       adsurl = {https://ui.adsabs.harvard.edu/abs/2015RAA....15.1095L}
}

@ARTICLE{EI19,
       author = {{Eilers}, Anna-Christina and {Hogg}, David W. and {Rix}, Hans-Walter and {Ness}, Melissa K.},
        title = "{The Circular Velocity Curve of the Milky Way from 5 to 25 kpc}",
      journal = {\apj},
         year = 2019,
        month = jan,
       volume = {871},
       number = {1},
          eid = {120},
        pages = {120},
          doi = {10.3847/1538-4357/aaf648},
archivePrefix = {arXiv},
       eprint = {1810.09466},
 primaryClass = {astro-ph.GA},
       adsurl = {https://ui.adsabs.harvard.edu/abs/2019ApJ...871..120E}
}

@ARTICLE{RI23,
       author = {{Ripepi}, V. and {Clementini}, G. and {Molinaro}, R. and {Leccia}, S. and {Plachy}, E. and {Moln{\'a}r}, L. and {Rimoldini}, L. and {Musella}, I. and {Marconi}, M. and {Garofalo}, A. and {Audard}, M. and {Holl}, B. and {Evans}, D.~W. and {Jevardat de Fombelle}, G. and {Lecoeur-Taibi}, I. and {Marchal}, O. and {Mowlavi}, N. and {Muraveva}, T. and {Nienartowicz}, K. and {Sartoretti}, P. and {Szabados}, L. and {Eyer}, L.},
        title = "{Gaia Data Release 3. Specific processing and validation of all sky RR Lyrae and Cepheid stars: The Cepheid sample}",
      journal = {\aap},
         year = 2023,
        month = jun,
       volume = {674},
          eid = {A17},
        pages = {A17},
          doi = {10.1051/0004-6361/202243990},
archivePrefix = {arXiv},
       eprint = {2206.06212},
 primaryClass = {astro-ph.SR},
       adsurl = {https://ui.adsabs.harvard.edu/abs/2023A&A...674A..17R}
}

@ARTICLE{CL19,
       author = {{Clementini}, G. and {Ripepi}, V. and {Molinaro}, R. and {Garofalo}, A. and {Muraveva}, T. and {Rimoldini}, L. and {Guy}, L.~P. and {Jevardat de Fombelle}, G. and {Nienartowicz}, K. and {Marchal}, O. and {Audard}, M. and {Holl}, B. and {Leccia}, S. and {Marconi}, M. and {Musella}, I. and {Mowlavi}, N. and {Lecoeur-Taibi}, I. and {Eyer}, L. and {De Ridder}, J. and {Regibo}, S. and {Sarro}, L.~M. and {Szabados}, L. and {Evans}, D.~W. and {Riello}, M.},
        title = "{Gaia Data Release 2. Specific characterisation and validation of all-sky Cepheids and RR Lyrae stars}",
      journal = {\aap},
         year = 2019,
        month = feb,
       volume = {622},
          eid = {A60},
        pages = {A60},
          doi = {10.1051/0004-6361/201833374},
archivePrefix = {arXiv},
       eprint = {1805.02079},
 primaryClass = {astro-ph.SR},
       adsurl = {https://ui.adsabs.harvard.edu/abs/2019A&A...622A..60C}
}

@ARTICLE{BU93,
       author = {{Butler}, R.~P.},
        title = "{Cepheid Velocity Curves from Lines of Different Excitation and Ionization. I. Observations}",
      journal = {\apj},
         year = 1993,
        month = sep,
       volume = {415},
        pages = {323},
          doi = {10.1086/173166},
       adsurl = {https://ui.adsabs.harvard.edu/abs/1993ApJ...415..323B}
}

@ARTICLE{CL23,
       author = {{Clementini}, G. and {Ripepi}, V. and {Garofalo}, A. and {Molinaro}, R. and {Muraveva}, T. and {Leccia}, S. and {Rimoldini}, L. and {Holl}, B. and {Jevardat de Fombelle}, G. and {Sartoretti}, P. and {Marchal}, O. and {Audard}, M. and {Nienartowicz}, K. and {Andrae}, R. and {Marconi}, M. and {Szabados}, L. and {Evans}, D.~W. and {Lecoeur-Taibi}, I. and {Mowlavi}, N. and {Musella}, I. and {Eyer}, L.},
        title = "{Gaia Data Release 3. Specific processing and validation of all-sky RR Lyrae and Cepheid stars: The RR Lyrae sample}",
      journal = {\aap},
         year = 2023,
        month = jun,
       volume = {674},
          eid = {A18},
        pages = {A18},
          doi = {10.1051/0004-6361/202243964},
archivePrefix = {arXiv},
       eprint = {2206.06278},
 primaryClass = {astro-ph.SR},
       adsurl = {https://ui.adsabs.harvard.edu/abs/2023A&A...674A..18C}
}

@ARTICLE{RI19,
       author = {{Ripepi}, V. and {Molinaro}, R. and {Musella}, I. and {Marconi}, M. and {Leccia}, S. and {Eyer}, L.},
        title = "{Reclassification of Cepheids in the Gaia Data Release 2. Period-luminosity and period-Wesenheit relations in the Gaia passbands}",
      journal = {\aap},
         year = 2019,
        month = may,
       volume = {625},
          eid = {A14},
        pages = {A14},
          doi = {10.1051/0004-6361/201834506},
archivePrefix = {arXiv},
       eprint = {1810.10486},
 primaryClass = {astro-ph.SR},
       adsurl = {https://ui.adsabs.harvard.edu/abs/2019A&A...625A..14R}
}

@ARTICLE{H24,
       author = {{Huang}, Yang and {Feng}, Qikang and {Khachaturyants}, Tigran and {Zhang}, Huawei and {Liu}, Jifeng and {Shen}, Juntai and {Beers}, Timothy C. and {Lu}, Youjun and {Wang}, Song and {Yuan}, Haibo},
        title = "{A slightly oblate dark matter halo revealed by a retrograde precessing Galactic disk warp}",
      journal = {Nature Astronomy},
         year = 2024,
        month = oct,
       volume = {8},
       number = {10},
        pages = {1294-1301},
          doi = {10.1038/s41550-024-02309-5},
archivePrefix = {arXiv},
       eprint = {2407.00319},
 primaryClass = {astro-ph.GA},
       adsurl = {https://ui.adsabs.harvard.edu/abs/2024NatAs...8.1294H}
}

@ARTICLE{JU08,
       author = {{Juri{\'c}}, Mario and {Ivezi{\'c}}, {\v{Z}}eljko and {Brooks}, Alyson and {Lupton}, Robert H. and {Schlegel}, David and {Finkbeiner}, Douglas and {Padmanabhan}, Nikhil and {Bond}, Nicholas and {Sesar}, Branimir and {Rockosi}, Constance M. and {Knapp}, Gillian R. and {Gunn}, James E. and {Sumi}, Takahiro and {Schneider}, Donald P. and {Barentine}, J.~C. and {Brewington}, Howard J. and {Brinkmann}, J. and {Fukugita}, Masataka and {Harvanek}, Michael and {Kleinman}, S.~J. and {Krzesinski}, Jurek and {Long}, Dan and {Neilsen}, Jr., Eric H. and {Nitta}, Atsuko and {Snedden}, Stephanie A. and {York}, Donald G.},
        title = "{The Milky Way Tomography with SDSS. I. Stellar Number Density Distribution}",
      journal = {\apj},
         year = 2008,
        month = feb,
       volume = {673},
       number = {2},
        pages = {864-914},
          doi = {10.1086/523619},
archivePrefix = {arXiv},
       eprint = {astro-ph/0510520},
 primaryClass = {astro-ph},
       adsurl = {https://ui.adsabs.harvard.edu/abs/2008ApJ...673..864J}
}

@ARTICLE{SH10,
       author = {{Sch{\"o}nrich}, Ralph and {Binney}, James and {Dehnen}, Walter},
        title = "{Local kinematics and the local standard of rest}",
      journal = {\mnras},
         year = 2010,
        month = apr,
       volume = {403},
       number = {4},
        pages = {1829-1833},
          doi = {10.1111/j.1365-2966.2010.16253.x},
archivePrefix = {arXiv},
       eprint = {0912.3693},
 primaryClass = {astro-ph.GA},
       adsurl = {https://ui.adsabs.harvard.edu/abs/2010MNRAS.403.1829S}
}

@ARTICLE{RE04,
       author = {{Reid}, M.~J. and {Brunthaler}, A.},
        title = "{The Proper Motion of Sagittarius A*. II. The Mass of Sagittarius A*}",
      journal = {\apj},
         year = 2004,
        month = dec,
       volume = {616},
       number = {2},
        pages = {872-884},
          doi = {10.1086/424960},
archivePrefix = {arXiv},
       eprint = {astro-ph/0408107},
 primaryClass = {astro-ph},
       adsurl = {https://ui.adsabs.harvard.edu/abs/2004ApJ...616..872R}
}

@ARTICLE{HU15,
       author = {{Huang}, Y. and {Liu}, X. -W. and {Yuan}, H. -B. and {Xiang}, M. -S. and {Huo}, Z. -Y. and {Chen}, B. -Q. and {Zhang}, Y. and {Hou}, Y. -H.},
        title = "{Determination of the local standard of rest using the LSS-GAC DR1}",
      journal = {\mnras},
         year = 2015,
        month = may,
       volume = {449},
       number = {1},
        pages = {162-174},
          doi = {10.1093/mnras/stv204},
archivePrefix = {arXiv},
       eprint = {1501.07095},
 primaryClass = {astro-ph.GA},
       adsurl = {https://ui.adsabs.harvard.edu/abs/2015MNRAS.449..162H}
}

@ARTICLE{WF21,
       author = {{Wang}, F. and {Zhang}, H. -W. and {Huang}, Y. and {Chen}, B. -Q. and {Wang}, H. -F. and {Wang}, C.},
        title = "{Local stellar kinematics and Oort constants from the LAMOST A-type stars}",
      journal = {\mnras},
         year = 2021,
        month = jun,
       volume = {504},
       number = {1},
        pages = {199-207},
          doi = {10.1093/mnras/stab848},
archivePrefix = {arXiv},
       eprint = {2103.10232},
 primaryClass = {astro-ph.GA},
       adsurl = {https://ui.adsabs.harvard.edu/abs/2021MNRAS.504..199W}
}

@BOOK{BT87,
       author = {{Binney}, James and {Tremaine}, Scott},
        title = "{Galactic dynamics}",
         year = 1987,
       adsurl = {https://ui.adsabs.harvard.edu/abs/1987gady.book.....B}
}

@ARTICLE{BV15,
       author = {{B{\"u}denbender}, Alex and {van de Ven}, Glenn and {Watkins}, Laura L.},
        title = "{The tilt of the velocity ellipsoid in the Milky Way disc}",
      journal = {\mnras},
         year = 2015,
        month = sep,
       volume = {452},
       number = {1},
        pages = {956-968},
          doi = {10.1093/mnras/stv1314},
archivePrefix = {arXiv},
       eprint = {1407.4808},
 primaryClass = {astro-ph.GA},
       adsurl = {https://ui.adsabs.harvard.edu/abs/2015MNRAS.452..956B}
}

@ARTICLE{OU24,
       author = {{Ou}, Xiaowei and {Eilers}, Anna-Christina and {Necib}, Lina and {Frebel}, Anna},
        title = "{The dark matter profile of the Milky Way inferred from its circular velocity curve}",
      journal = {\mnras},
         year = 2024,
        month = feb,
       volume = {528},
       number = {1},
        pages = {693-710},
          doi = {10.1093/mnras/stae034},
archivePrefix = {arXiv},
       eprint = {2303.12838},
 primaryClass = {astro-ph.GA},
       adsurl = {https://ui.adsabs.harvard.edu/abs/2024MNRAS.528..693O}
}

@ARTICLE{BJ05,
       author = {{Benjamin}, R.~A. and {Churchwell}, E. and {Babler}, B.~L. and {Indebetouw}, R. and {Meade}, M.~R. and {Whitney}, B.~A. and {Watson}, C. and {Wolfire}, M.~G. and {Wolff}, M.~J. and {Ignace}, R. and {Bania}, T.~M. and {Bracker}, S. and {Clemens}, D.~P. and {Chomiuk}, L. and {Cohen}, M. and {Dickey}, J.~M. and {Jackson}, J.~M. and {Kobulnicky}, H.~A. and {Mercer}, E.~P. and {Mathis}, J.~S. and {Stolovy}, S.~R. and {Uzpen}, B.},
        title = "{First GLIMPSE Results on the Stellar Structure of the Galaxy}",
      journal = {\apjl},
         year = 2005,
        month = sep,
       volume = {630},
       number = {2},
        pages = {L149-L152},
          doi = {10.1086/491785},
archivePrefix = {arXiv},
       eprint = {astro-ph/0508325},
 primaryClass = {astro-ph},
       adsurl = {https://ui.adsabs.harvard.edu/abs/2005ApJ...630L.149B}
}

@ARTICLE{BV12a,
       author = {{Bovy}, Jo and {Rix}, Hans-Walter and {Liu}, Chao and {Hogg}, David W. and {Beers}, Timothy C. and {Lee}, Young Sun},
        title = "{The Spatial Structure of Mono-abundance Sub-populations of the Milky Way Disk}",
      journal = {\apj},
         year = 2012,
        month = jul,
       volume = {753},
       number = {2},
          eid = {148},
        pages = {148},
          doi = {10.1088/0004-637X/753/2/148},
archivePrefix = {arXiv},
       eprint = {1111.1724},
 primaryClass = {astro-ph.GA},
       adsurl = {https://ui.adsabs.harvard.edu/abs/2012ApJ...753..148B}
}

@ARTICLE{BH16,
       author = {{Bland-Hawthorn}, Joss and {Gerhard}, Ortwin},
        title = "{The Galaxy in Context: Structural, Kinematic, and Integrated Properties}",
      journal = {\araa},
         year = 2016,
        month = sep,
       volume = {54},
        pages = {529-596},
          doi = {10.1146/annurev-astro-081915-023441},
archivePrefix = {arXiv},
       eprint = {1602.07702},
 primaryClass = {astro-ph.GA},
       adsurl = {https://ui.adsabs.harvard.edu/abs/2016ARA&A..54..529B}
}

@ARTICLE{CH17,
       author = {{Chen}, B. -Q. and {Liu}, X. -W. and {Yuan}, H. -B. and {Robin}, A.~C. and {Huang}, Y. and {Xiang}, M. -S. and {Wang}, C. and {Ren}, J. -J. and {Tian}, Z. -J. and {Zhang}, H. -W.},
        title = "{Constraining the Galactic structure parameters with the XSTPS-GAC and SDSS photometric surveys}",
      journal = {\mnras},
         year = 2017,
        month = jan,
       volume = {464},
       number = {3},
        pages = {2545-2556},
          doi = {10.1093/mnras/stw2497},
archivePrefix = {arXiv},
       eprint = {1609.08838},
 primaryClass = {astro-ph.GA},
       adsurl = {https://ui.adsabs.harvard.edu/abs/2017MNRAS.464.2545C}
}

@ARTICLE{WN17,
       author = {{Wan}, Jun-Chen and {Liu}, Chao and {Deng}, Li-Cai},
        title = "{Red Clump Stars from LAMOST II: the outer disc of the Milky Way}",
      journal = {Research in Astronomy and Astrophysics},
         year = 2017,
        month = aug,
       volume = {17},
       number = {8},
          eid = {079},
        pages = {079},
          doi = {10.1088/1674-4527/17/8/79},
archivePrefix = {arXiv},
       eprint = {1704.04369},
 primaryClass = {astro-ph.GA},
       adsurl = {https://ui.adsabs.harvard.edu/abs/2017RAA....17...79W}
}

\appendix
\renewcommand{\thefigure}{A\arabic{figure}}

\section*{APPENDIX: RC Measurement using distances from SK25}

\begin{figure}
    \begin{center}
    \includegraphics[width=1.0\linewidth]{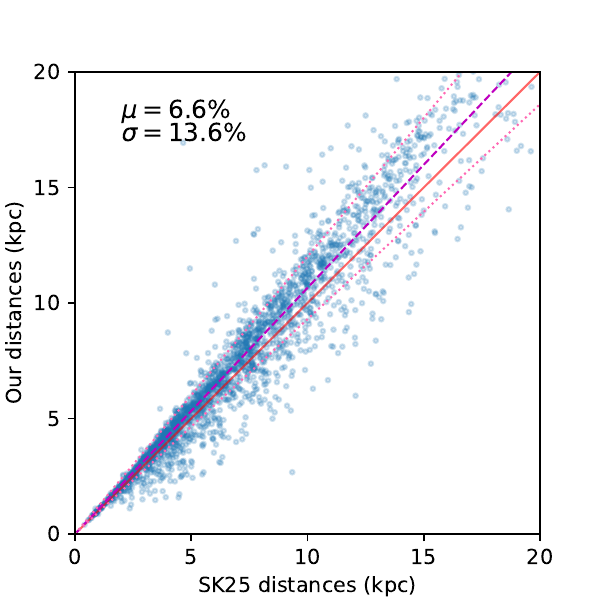}
    \end{center}
    \caption{Comparison between our distances and those from SK25. The mean ($\mu$) and standard deviation ($\sigma$) of the relative distance difference ($\Delta d/d_{\mathrm{SK25}}$) are shown in the upper-left corner. The magenta dashed line represents $d=(1+\mu)d_{\mathrm{SK25}}$, while the two pink dotted lines indicate $d=(1+\mu+\sigma)d_{\mathrm{SK25}}$ and $d=(1+\mu-\sigma)d_{\mathrm{SK25}}$, respectively.}
\end{figure}

\begin{figure}
    \begin{center}
    \includegraphics[width=1.0\linewidth]{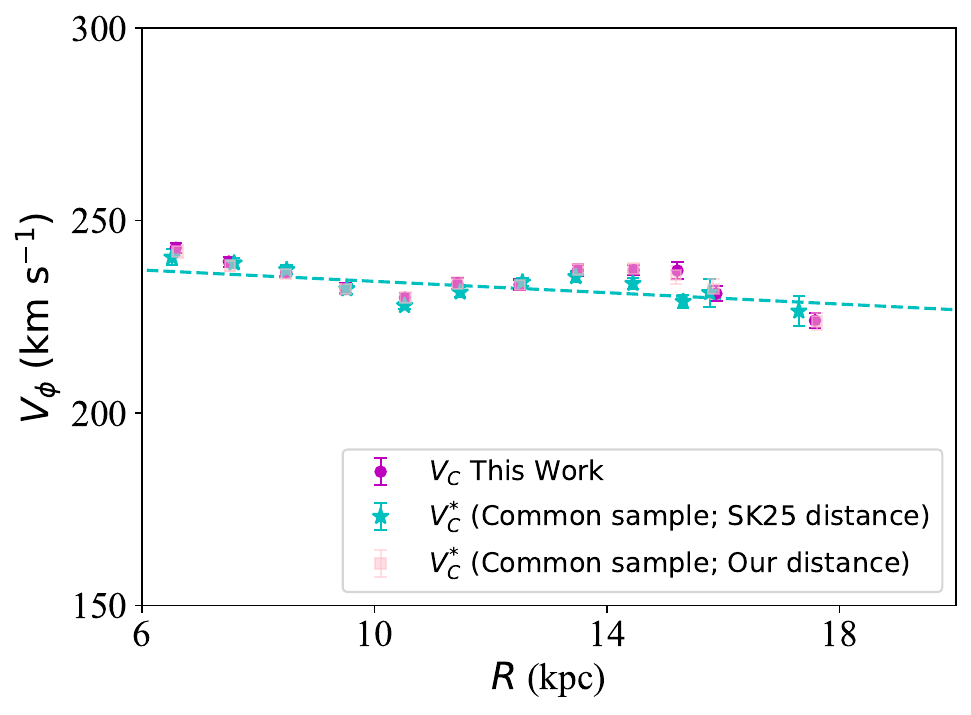}
    \end{center}
    \caption{Comparison of the test RCs and our main result. The magenta dots denote our RC result, while the cyan and pink dots represent the test RCs derived using distances from SK25 and our distances, respectively.}
\end{figure}

To assess whether small differences in distance estimates, such as those between our measurements and those reported by SK25, could lead to significant discrepancies in the RC determination, we conduct a test by repeating the RC measurement procedure with different distances.

Specifically, we first cross-match our sample with that of SK25 and apply the same selection criteria described in Section 2 to the common sample, leaving 760 stars. Using this common sample, for which both SK25 distances and our distances are available, we perform the RC measurements both with SK25 distances and with our distances.

Figure A2 presents the test RCs obtained from the common sample, alongside our RC result (derived in Section 3.2). The test RC derived using SK25 distances (cyan dots) shows excellent agreement with our RC result (magenta dots), reproducing both the overall decreasing trend and the dip–bump feature. Also, no clear systematic offsets are observed between them. In nearly all radial bins, the difference between the test RC from SK25 distances and either our RC result or the test RC derived from our own distances (pink dots) is less than $2.5\ \mathrm{km\ s^{-1}}$. This comparison demonstrates that minor discrepancies between our distances and those from SK25 do not significantly affect the RC measurement.

\bsp	
\label{lastpage}
\end{document}